\documentclass[aps,nofootinbib,floatfix,showpacs,preprintnumbers]{revtex4} 
\usepackage{graphicx}
\input epsf
\newcommand{\sfig}[2]{
\includegraphics[width=#2]{#1}
        }

\newcommand{\Sfig}[2]{
    \begin{figure}[thbp]
    \sfig{#1.eps}{.7\columnwidth}
    \caption{{\small #2}}
    \label{fig:#1}
    \end{figure}
}\newcommand{\Sfigc}[3]{
    \begin{figure}[thbp]
    \sfig{#2.eps}{#1\columnwidth}
    \caption{{\small #3}}
    \label{fig:#2}
    \end{figure}
}
\newcommand{\Mapfig}[3]{
    \begin{figure}[thbp]
    \includegraphics[width=#1\columnwidth, bb = 0 250 595 842]{#2.ps}
    \caption{{\small #3}}
    \label{fig:#2}
    \end{figure}
}

\newcommand{\rf}[1]{\ref{fig:#1}}

\def\lsim{\mathrel{\raise.3ex\hbox{$<$\kern-.75em\lower1ex\hbox{$\sim$}}}}
\def\gsim{\mathrel{\raise.3ex\hbox{$>$\kern-.75em\lower1ex\hbox{$\sim$}}}}

\def\cmm2{{\,\rm cm^{-2}}}
\def\cm2{{\,{\rm cm}^2}}
\def\cmm3{{\,{\rm cm}^{-3}}}
\def\gcmm3{{\,{\rm g\,cm^{-3}}}}

\def\fun#1#2{\lower3.6pt\vbox{\baselineskip0pt\lineskip.9pt
  \ialign{$\mathsurround=0pt#1\hfil##\hfil$\crcr#2\crcr\sim\crcr}}}

\def\be{\begin{equation}}
\def\ee{\end{equation}}
\def\bea{\begin{eqnarray}}
\def\eea{\end{eqnarray}}

\newcommand{\ec}[1]{Eq.~(\ref{eq:#1})}

\newcommand{\eql}[1]{\label{eq:#1}}

\newcommand{\ppp}{f_{\rm SUSY}}

\begin{document}

\title{Constraining Dark Matter in Galactic Substructure}

\author{Eric J. Baxter$^1$, Scott Dodelson$^{1,2,3}$, Savvas M. Koushiappas$^4$, Louis E. Strigari$^5$
}

\affiliation{$^1$Department of Astronomy \& Astrophysics, The
University of Chicago, Chicago, IL~~60637}
\affiliation{$^2$Center for Particle Astrophysics, Fermi National
Accelerator Laboratory, Batavia, IL~~60510}
\affiliation{$^3$Kavli Institute for Cosmological Physics, Chicago, IL~~60637}
\affiliation{$^4$Department of Physics, Brown University, Providence, RI~~02912}
\affiliation{$^5$Kavli Institute for Particle Astrophysics and Cosmology, Stanford University, Stanford, CA~~94305}
\date{\today}
\begin{abstract}
Detecting the dark matter annihilation signal from Galactic
substructure, or subhalos, is an important challenge for high-energy
gamma-ray experiments. In this paper we discuss detection prospects by
combining two different aspects of the gamma-ray signal: the angular
distribution and the photon counts probability distribution function
(PDF).  The true PDF from subhalos has been shown recently (by Lee et
al.) to deviate from Poisson; we extend this analysis and derive the
signal PDF from a detailed $\Lambda$CDM-based model for the properties
of subhalos.  We combine our PDF with a model for Galactic and
extra-Galactic diffuse gamma-ray emission to obtain an estimator and
projected error on dark matter particle properties (mass and
annihilation cross section) using the Fermi Gamma-Ray Space
Telescope. We compare the estimator obtained from the true PDF to that
obtained from the simpler Poisson analysis. We find that, although
both estimators are unbaised in the presence of backgrounds, the error
on dark matter properties derived from the true PDF is
$\sim 50\%$ smaller than when utilizing the Poisson-based analysis.
\end{abstract}
\maketitle

\section{Introduction}

A wide variety of evidence points to the existence of nonbaryonic dark
matter~\cite{Bertone:2004pz}. There are three ways of directly
confirming this hypothesis: producing dark matter or its cousins in an
accelerator \cite{Birkedal:2004, Feng:2006}, directly detecting dark
matter particles impinging on the Earth in underground detectors
\cite{Drukier:1984,Goodman:1985}, and indirectly detecting dark matter
by observing the products of an annihilation of two dark matter
particles in space \cite{Silk:1984,Gunn:1978,Golubkov:1999}.  The
current excitement in the field stems from the coincidental maturity
of all three of these techniques. The Large Hadron Collider began
operations in 2009, a number of direct detection experiments have
proven their ability to scale up to the one-ton level, and there are
several experiments (Fermi Gamma-Ray Satellite
Telescope~\cite{Baltz:2008wd}, Atmospheric Cerenkov
Telescopes~\cite{Hui:2008tt}, PAMELA~\cite{Adriani:2008zr}, Ice
Cube~\cite{Resconi:2008fe}) poised to detect the indirect signal.

For gamma-ray experiments a key challenge is to extract the dark
matter signal in the presence of emission from point sources, such as
pulsars and AGN, and diffuse sources such as cosmic rays. One way to
discriminate photons produced by a given dark matter source from the
above backgrounds is to measure the energy spectrum.  Photons generated by
the annihilation of standard thermally-produced particle dark matter
have a spectrum characteristic of quark production and hadronization\cite{Bergstrom:1997fj,Fornengo:2004kj}, 
distinguishing them from the
typical power law-like behavior of other
sources~\cite{Baltz:2006sv,Venters:2007rn}.  A second discriminant is
the angular
distribution~\cite{Hooper:2007be,Cuoco:2007sh,Dodelson:2007gd,Kuhlen:2007wv,Kuhlen:2008aw}.
The angular distribution of photons produced in dark matter
annihilations results from the variation of the dark matter density
profile as a function of $\psi$, the angle between the incoming
direction and the line connecting us to the Galactic center.  By
contrast, the extra-Galactic
background~\cite{Stecker:2001dk,Dermer:2007fg} is more or less
isotropic, and the diffuse Galactic background is predominantly
confined to the Galactic disk.

Recently, several
groups~\cite{Lee:2008fm,Dodelson:2009ih,SiegalGaskins:2009ux} have
explored the possibility of indirect detection in the Milky Way halo
by adding another discriminant, the probability distribution function
(PDF). In their recent analysis Lee et al.~\cite{Lee:2008fm} have
determined the PDF of photons produced by dark matter annihilations in
dark matter substructure (subhalos) in our Galaxy, and they have shown
that this PDF is clearly distinct from a Poisson distribution. In
particular, for a given pixel observed by, e.g., the Fermi telescope,
there is an unusually large probability (unusual compared with Poisson
expectation for the same mean number of counts) of observing multiple
counts from the population of subhalos along the line-of-sight.

Here we test the idea of using the PDF, together with a
$\Lambda$CDM-based model for the scatter in subhalo properties, to
extract the dark matter signal in the Fermi experiment. In this work
we are in particular interested in answering the following questions:
\begin{itemize}
\item Can the PDF -- if known -- be used as an effective tool to
  extract the dark matter signal?
\item Will Fermi have the statistical reach to probe a velocity-weighted annihilation
  cross section of $3\times 10^{-26}$ cm$^3$ sec$^{-1}$, the canonical
  value for a thermally produced dark matter candidate?
\item Does one need to know the PDF in order to analyze an experiment?
  I.e., if one incorrectly assumes a $\chi^2$ distribution, will s/he
  be led to incorrect conclusions about the parameters under
  consideration?
\end{itemize}
For concreteness, throughout, we make predictions and projections for one year of Fermi data.

The layout of this paper is as follows.  \S{\ref{sec:model}} describes
the 3-component model of subhalo, diffuse Galactic and extra-Galactic
emission that we use.  Simulated maps of these components are produced
in \S{\ref{sec:map}}. We then analyze the maps two different ways in
\S{\ref{sec:chi}}: with a standard $\chi^2$ analysis and with the
exact likelihood. The former does not use the information contained in
the PDF, while the latter analysis does use this information.  Our
conclusions are presented in \S{\ref{sec:con}}.

\section{The Model}\label{sec:model}

We assume a three-component model for the diffuse gamma-ray
background: annihilation radiation from dark matter subhalos, Galactic
emission, and extra-Galactic emission. This simplified model neglects
other contributions to the gamma-ray background, including point
sources -- both Galactic and extra-Galactic -- which we assume can be
identified and removed. We also neglect other dark matter sources,
including diffuse emission from the Milky Way halo and from
cosmological halos, as we are in particular concerned with isolating
the subhalo contribution. Given the latitudes that we consider for our
analysis this model is appropriate~\cite{Springel:2008zz}.  As we
argue here, even our simplified model represents an improvement in our
understanding of diffuse emission from subhalos and our ability to
extract it using gamma-ray data. In the context of the larger goal of
detecting dark matter our assumptions may be viewed as conservative,
as we are neglecting several possible sources of signal.

Following~\citet{Lee:2008fm}, we write the probability of
obtaining $C_i$ counts in bin $i$ which is an angle $\psi_i$ away from
the Galactic center as
\begin{equation}
\label{eq:Pci}
P(C_i) = \int dF P_{sh}(F;\psi_i) \mathcal{P}[E_iF+C^{\rm gal}_i+C^{\rm eg}_i;C_i], 
\eql{prob}
\end{equation}
where $P_{sh}(F;\psi_i)$ is the probability of subhalos producing a
flux $F$ which depends on $\psi_i$ in the pixel; $\mathcal{P}$ is the
Poisson probability for obtaining $C_i$ counts if the mean number of
counts is equal to $F$ multiplied by the exposure of the pixel in the
experiment, $E_i$, plus the counts expected from the two background
sources, $C^{\rm gal}_i$ and $C^{\rm eg}_i$. We are implicitly
assuming here that the PDF's of both background components -- Galactic
and extra-Galactic -- are Poisson, as opposed to the PDF of the
subhalo contribution which is captured in $P_{sh}$. This is the best
one could hope for when examining the utility of the PDF; if the PDF
turns out not to matter much in our analysis, then this will be a
robust conclusion. In the rest of this section, we describe the
details of this model, now specified by $P_{sh}(F;\psi_i)$ and the
expected number of counts due to backgrounds $C^{\rm gal}_i$ and
$C^{\rm eg}_i$.

\subsection{The Signal: Emission from Subhalos}


In order to calculate the counts probability distribution function
given in Eq.~\ref{eq:Pci}, we must estimate the flux probability
distribution $P_{sh}(F;\psi_i)$, which depends on a description of the
abundance and properties of all subhalos along the line of sight.
Following~\citet{Lee:2008fm}, we first calculate $P_1(F;\psi_i)$, the
probability of observing a flux $F$ from a \textit{single} subhalo at
angle $\psi_i$ from the Galactic center:
\begin{equation} 
\label{eq:P1}
P_{1}(F;\psi_i) \propto \Theta(F_{\mathrm{max}} - F
)\int_0^{\ell_{\mathrm {max} }} d\ell \int dL_{sh} \,
P(L_{sh},\ell,\psi_i) \delta \left( F - \frac{ L_{sh}}{4 \pi \ell^2}
\right) .
\end{equation}
Here, $ P(L_{sh},\ell,\psi_i)$ is the probability of finding a subhalo
emitting luminosity $L_{sh}$ at a distance $\ell$ from us at an angle
$\psi_i$ from the Galactic center. The step function limits the flux
to be less than $F_{\mathrm{max}}$ since sources with larger fluxes
will be identified as resolved point sources. Although the resolved
flux limit of Fermi depends on
energy\footnote{http://www-glast.slac.stanford.edu/software/IS/.}, for
concreteness we choose a simple threshold of $F_{\mathrm{max}}=
10^{-9}\mathrm{cm}^{-2}\mathrm{s}^{-1}$.  The line-of-sight integral
extends out to $\ell_{\rm max}$, which is determined by the assumed
extent of the dark matter halo. The probability
$P(L_{sh},\ell,\psi_i)$ can be broken up into a convolution of the
well-studied mass function with the conditional luminosity function:
\begin{equation}
P(L_{sh},\ell,\psi_i) d\ell
\propto \ell^2 d\ell \int_{M_{\mathrm{min}}}^{M_{\mathrm{max}}} dM 
\frac{dN\left[r(\ell,\psi_i)\right]}{dM\,dV}
P[L_{sh} \,\vert\,M,r(\ell, \psi_i)], 
\end{equation}
with $r(\ell, \psi_i) = \sqrt{\ell^2 + d_\odot^2 - 2 \ell d_\odot \cos
  \psi_i}$ where $d_\odot= 8.5 \, \mathrm{kpc}$ is the Galactocentric
distance of the Sun.  The assumption that the dark matter halo extends
out to $R_G=220$ kpc leads to $\ell_{\mathrm{max}} = d_\odot \left[
  \cos \psi_i + \sqrt{-\sin^2 \psi_i + (R_G / d_\odot)^2}\right]$.
The lower limit on the mass integral, $M_{\mathrm{min}}$, is
determined by the cutoff scale of the subhalo mass function in the
Milky Way halo.  Supersymmetric models with WIMP dark matter
candidates typically have a cutoff scale in the dark matter power
spectrum in the range $M_{\mathrm{min}} \sim 10^{-6}-10^0 M_\odot$
\cite{HSS01,CKZ02,Profumo:2006bv,Schmid:1998mx,Green:2003un,Green:2005fa,
  LZ05,Martinez:2009jh}; motivated by these models, for all results
here we will adopt a value of $M_{\mathrm{min}}=0.01 M_\odot$.  We
discuss the impact of varying $M_{\rm{min}}$ about this fiducial value
below.  The upper limit on the halo mass (which is not particularly
relevant since the mass function falls off fairly steeply) is taken to
be $10^{10} M_\odot$.  With this information, Eq.~\ref{eq:P1} can now
be written as
\begin{equation} 
P_{1}(F;\psi_i) \propto \Theta(F_{\mathrm{max}} - F )
\int_0^{\ell_{\mathrm {max} }}\,d\ell \, \ell^4
\int_{M_{\mathrm{min}}}^{M_{\mathrm{ max}}} \, dM \frac{dN[r(\ell,
\psi_i)]}{dMdV} P[L_{sh} = 4 \pi \ell^2 F\,\vert\,M,r(\ell, \psi_i)].
\end{equation}

To complete this calculation, we need the mass function and
conditional luminosity function.  In~\citet{Lee:2008fm}, it was
assumed that there is a one-to-one mapping between subhalo luminosity
and the mass of a subhalo, namely $L_{sh} \propto M_{sh}$. For our
analysis we determine the $L_{sh} - M_{sh}$ relation using the
properties of simulated subhalos in a $\Lambda$CDM cosmology
~\cite{Koushiappas:2010}. The properties of subhalos, including those
that will be relevant for us such as the spatial distribution and the
assigned gamma-ray luminosity, reflect the underlying process of
non-linear structure growth. The complex interplay between formation
redshift, time of accretion to the parent halo, and orbital and tidal
evolution sets the characteristics of the luminosity-mass relationship
of subhalos, as well as the radial distribution (see
\cite{Koushiappas:2010}). As a result of this process, subhalos with
similar mass and Galactocentric radius will have a spread in their
gamma-ray luminosities.

We include this non-zero scatter by using the conditional luminosity
distribution found in~\cite{Koushiappas:2010}, 
\begin{equation} 
P(\ln L_{sh}\,\vert\,M,r) = \frac{1}{\sqrt{2 \pi}} \frac{1}{\sigma} \exp \left[ -
  \frac{ [ \ln L_{sh} - \langle \ln L_{sh} \rangle ]^2 }{2 \sigma^2}
\right]. 
\end{equation}
For a dark matter halo with a concentration of approximately $c
\approx 10$ (model $C_0$ in~\cite{Koushiappas:2010}), the mean
luminosity $\langle L_{sh} \rangle$, as well as the spread about the
mean luminosity $\sigma$ depend on subhalo mass and Galactocentric
radius via
\begin{equation}
\langle \ln ( L_{sh} / {\mathrm s}^{-1}) \rangle = 77.4 + 0.87
\ln(M/10^5M_\odot) - 0.23 \ln(r/50{\mathrm {kpc}}) + \ln\left(
\frac{\ppp}{10^{-28}\,{\rm cm}^3\,{\rm s}^{-1}\,{\rm GeV}^{-2}}
\right)\eql{lum}
\end{equation}
\begin{equation} 
\sigma = 0.74 - 0.0030 \ln(M/10^5M_\odot) - 0.011 \ln( r/50 {\mathrm
  {kpc}}).
\eql{sigma}
\end{equation}
The quantity $\ppp$ is the particle physics parameter\footnote{We use the $\ppp$ to conform to the literature, but nothing in our analysis depends on supersymmetry; all that matters is the combination of cross section, mass, and $N_\gamma$ folded into $\ppp$.} governing the
emission rate,
\begin{equation} 
\ppp = N_\gamma \frac{\langle \sigma v \rangle}{m_\chi^2}.
\end{equation}  
Here the mass of the dark matter particle is $m_\chi$, $\langle \sigma v \rangle$ is the
thermally averaged annihilation cross section times the velocity, and
$N_\gamma$ is the number of photons above 1 GeV emitted in the annihilation of a
single dark matter pair.  A thermally averaged cross section of
$\langle \sigma v\rangle=3\times 10^{-26}$ cm$^3$ s$^{-1}$ leads to
the correct thermal abundance of dark matter today so that our
fiducial value of $\ppp = 10^{-28}\,{\rm cm}^3\,{\rm
s}^{-1}\,{\rm GeV}^{-2}$ is easily accommodated in supersymmetric
models~\cite{Bergstrom:2001jj,Fornengo:2004kj}.

Thus the mean luminosity in \ec{lum} differs from that
of~\citet{Lee:2008fm} in several ways: it scales with mass as $L_{sh}\propto M^{0.87}$, in agreement with simple
analytic estimates~\cite{Strigari:2006rd}, as well as numerical
simulation results~\cite{Diemand:2006ik,Kuhlen:2008aw}). Furthermore,
the luminosity depends on the radial position of subhalos
($L_{sh} \propto r^{-0.23}$), and we also include a non-zero scatter
(\ec{sigma}) about the mean value of the luminosity, a scatter which depends
on the mass and the Galactocentric radius of subhalos.

Numerical simulations predict a mass function of the form 
\begin{equation}
\label{eq:massfunc}
dN(r)/dMdV = A \frac{(M/M_\odot)^{-\beta}} {\tilde{r} ( 1 + \tilde{r})^2},
\end{equation}
with $\beta \approx 1.9$~~\cite{Springel:2008cc}. Here the radial
dependence is through $\tilde{r} = r / r_s$, where $r_s$ is the scale
radius of the Milky Way halo ($r_s \approx 21 {\rm kpc}$).  We
normalize the mass function by utilizing the numerical result that
roughly 10\% of the mass of the Galactic halo ($M_G = 1.2 \times
10^{12} M_\odot$) is in subhalos of mass greater than $\sim 10^7
M_\odot$. With this assumption, the normalization constant is $A
\approx 1.2 \times 10^4 M_\odot^{-1} {\mathrm {kpc}^{-3}}$.
Simulations also suggest that the halo distribution may be less cuspy
near the center than the dark matter profile, and may depend on the
mass of the subhalo~\cite{D'Onghia:2009pz}. This may have implications
on the expected annihilation signal from substructure as the overall
number of counts along a particular line of sight will be lower than
expected (especially if most of the signal arrives from nearby
objects). Nevertheless, given the current uncertainties of the level
of this effect, we do not include a core in the distribution of
subhalos in this study, but we emphasize that the issue of
substructure depletion in the inner regions of the Galaxy must be
addressed in detail in future numerical simulations.

\Sfig{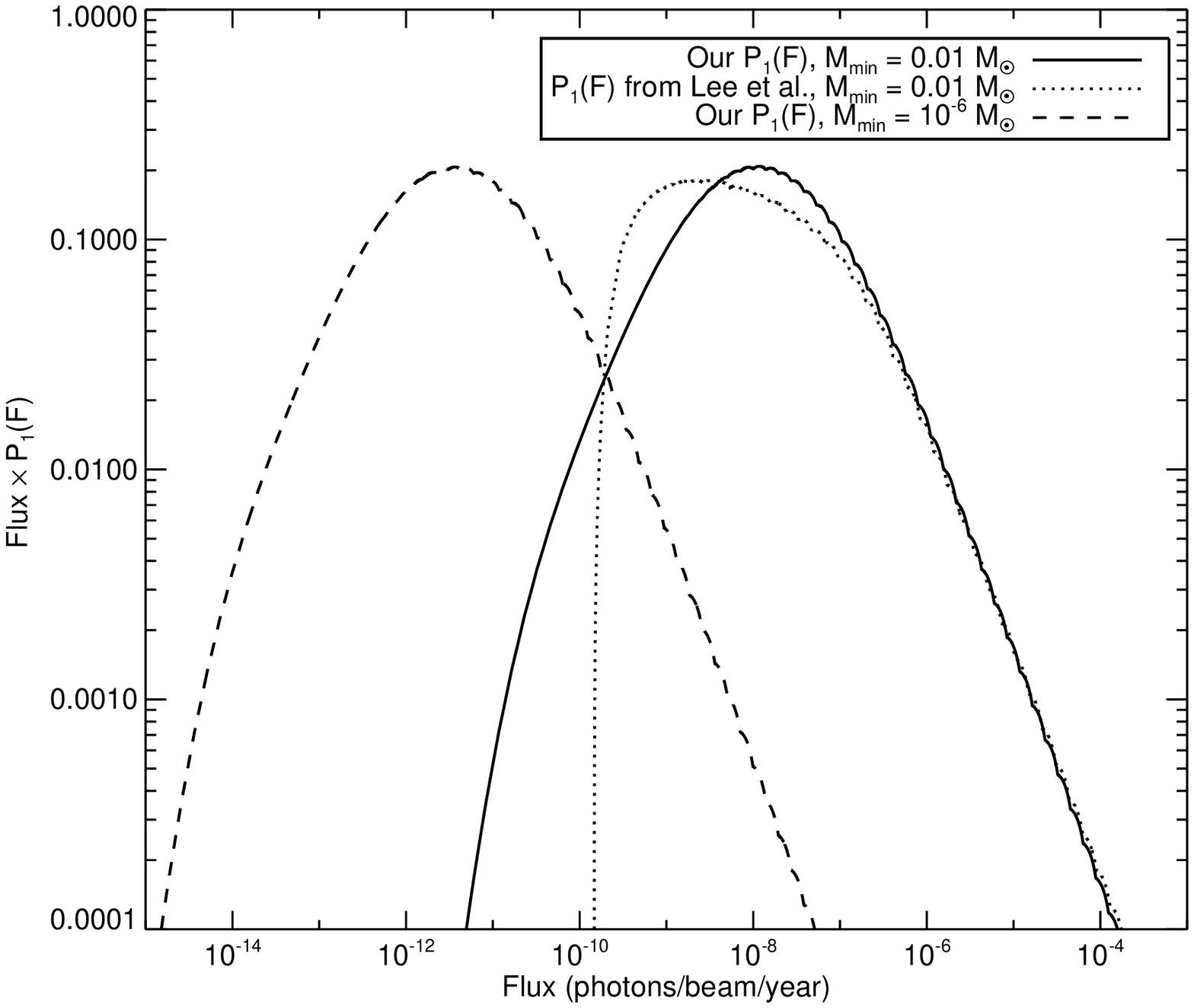}{Probability $P_1(F,\psi_i=40^\circ)$ of observing flux $F$
  from a single halo in a given square degree pixel.  We measure flux
  in units of photons/beam/year, where a `beam' corresponds to the
  approximate effective area of the Fermi telescope, $A \sim 2000 \rm{
    cm}^2$.  The solid curve uses luminosity and mass functions from
  this paper with $M_{\rm{min}} = 0.01 M_{\odot}$, while the dashed
  curve uses the same functions with $M_{\rm{min}} = 10^{-6}
  M_{\odot}$.  The dotted curve shows $P_1$ from
  Ref.~\cite{Lee:2008fm} with $M_{\rm{min}} = 0.01 M_{\odot}$,
  re-scaled to have the same mean flux as our $P_1$ for the purpose of
  comparison.}

With the above ingredients we construct the probability of observing a
single subhalo with flux $F$ in pixel $i$, $P_1(F,\psi_i)$, shown in
Fig.~\rf{f1} for $\psi_i=40^\circ$.  In generating this figure, we
have used flux units of photons/beam/year, where the beam corresponds
to the detector area of the Fermi telescope, $A \sim 2000\rm{ cm}^2$
(the true effective area of Fermi is energy dependent so this value is
only approximate).  Of particular note in this figure is the smoother
fall off at low flux in our model relative to the model
of~\citet{Lee:2008fm} (for the purpose of comparison we have scaled
the~\citet{Lee:2008fm} model so that it predicts the same mean flux as
our model).  As both of these models assume a sharp mass cut-off at
the low end, the difference in fall-off at low flux follows directly
from the scatter in luminosity for a given mass.  When the low end
mass cut-off, $M_{\rm{min}}$, is changed, we see from Fig.~\rf{f1}
that the mean flux per subhalo decreases but that the shape of
$P_1(F)$ remains essentially unchanged.

Also, note that the PDF's for both models are very similar at the high
flux end as a result of competing differences between the mass functions
and the mass-luminosity relations of the two models.  For a mass
function that goes as $dN/dM \propto M^{-\beta}$ and a mass-luminosity
relation such that $L_{sh} \propto M^{\alpha}$, it can be shown that
for large values of $F$, $P_{1}(F) \propto F^{\gamma}$ with $\gamma =
(1-\beta)/\alpha -1$.  For our model we have $\gamma = (1-1.9)/0.87 -1
= -2.03$ while for the~\citet{Lee:2008fm} model they have $\gamma =
(1-2)/1 - 1 = -2$.  Thus, in both models $P_{1}(F)$ is approximately
proportional to $F^{-2}$ for large $F$.  Physically, our less-steep
mass function means that we have more high mass (and thus high
luminosity) subhalos than the~\citet{Lee:2008fm} model, but our
less-steep luminosity function means that these subhalos are not as
bright.  The end result is that on the high flux end both models are
very similar.


We use $P_{1}(F;\psi_i)$ to determine $P_{sh}(F;\psi_i)$, the probability
of observing a total flux from multiple subhalos at angle $\psi_i$ from the
Galactic center.  The two functions are related by
\begin{equation}
\label{eq:Pf}
P_{sh}(F;\psi_i) =
\mathcal{F}^{-1}\left\{e^{\mu(\psi_i)\left(\mathcal{F}\left\{P_{1}(F;\psi_i)\right\}-1
\right)}\right\},
\end{equation}
where $\mathcal{F}$ indicates a Fourier transform with respect to $F$
and $\mu$ is the mean number of subhalos in a given pixel:
\begin{equation}
\mu(\psi_i) = \Omega_{pixel} \int d{\ell} \ell^2 \int dM
\frac{dN\left[r(\ell,\psi_i)\right]}{dMdV}.
\end{equation}
$\Omega_{pixel}$ is the solid angle of a single pixel, taken here
to be one square degree.
Eq.~\ref{eq:Pf} can be derived by assuming that the number of subhalos
contributing to the photon counts in a single pixel is a Poisson
random variable with mean $\mu$ and that each subhalo emits a flux $F$
with probability $P_1(F;\psi_i)$.  A detailed derivation of
Eq.~\ref{eq:Pf} is presented in the appendix of~\cite{Lee:2008fm}.

Finally, given $P_{sh}(F;\psi_i)$ we can construct $P(C_i)$, the
probability of getting $C$ counts in pixel $i$ by applying \ec{prob}.
Fig.~\rf{f2} shows the calculated $P(C_i)$ for the dark matter
signal for one year of observation by the Fermi Telescope.  Note that
a Poisson distribution with the same mean number of counts has a
significantly smaller probability of producing high-count pixels than
the true $P(C_i)$.  Also, note that despite the differences between
our model and that of \citet{Lee:2008fm}, both produce PDFs that
are very similar.  Apparently, the differences between the two models
are washed out through the transition to $P_{sh}(F;\psi_i)$ and the
subsequent discretization to produce $P(C_i)$.  The similarity between
the two models is encouraging: it suggests that the form of $P(C_i)$
is somewhat independent of the many assumptions that go into such
models (e.g. the mass function, the luminosity function, $M_{\rm{min}}$,
etc.), thus making our conclusions more robust.


In Table~\ref{table:counts} we show the expected number of counts for
our fiducial model, as well as four other models where we vary the
cutoff scale of the mass function and the concentration (and
substructure mass fraction) of the host Milky Way halo.  The effect of
the subhalo mass function cutoff on the photon counts is due to the
fact that the luminosity increases with mass at a slower rate than the
rate at which the abundance is increasing with mass - {\it numerous}
small (and faint) subhalos yield a higher flux than {\it few} large
and bright subhalos.  The effect is not very large, as decreasing
$M_{\rm{min}}$ by four orders of magnitude results in only a factor
of $\sim4.5$ increase in the photon counts.  Still, understanding the
low-mass cutoff scale of the subhalo mass function is important in any
future interpretation of $\gamma$-ray data.

The high and low concentration models in Table~\ref{table:counts}
refer to models $C_+$ and $C_-$ respectively in
\cite{Koushiappas:2010}. They represent host Milky Way halos with high
($c \ge 13$) and low ($c \le 7$) concentrations. The luminosity PDF is
a weak function of concentration, except perhaps in the very inner
regions of the halo.  The normalization of the subhalo mass function,
however, depends somewhat strongly on the host concentration. High concentration
host halos have a lower normalization of substructure $f \approx
0.08$ (where $f$ is the mass fraction of subhalos relative to the
total halo mass) relative to low concentration halos which have a
higher normalization of substructure $ f \approx 0.3$. This is an
outcome of hierarchical structure formation. High concentration host
halos were formed earlier and therefore their constituent subhalos
evolved for a longer period of time in the presence of the tidal field
of the host, thus the subhalo survival rate is lower than in the low
concentration (recently formed) hosts.  As can be seen from
Table~\ref{table:counts}, varying the host concentration changes the
total photon counts by at most about $60\%$.

\begin{table}[htb]\footnotesize
\begin{center}
\begin{tabular}{c|c|c}
\hline
Model & Mean Counts  & Approximate \\
 & at $\psi = 40^{\circ}$ & Total Counts \\ 
\hline
Fiducial & 0.83 & 6600 \\
Fiducial with $M_{\mathrm{min}} = 10^{-6} M_{\odot}$  & 1.36 & 29800 \\
Fiducial with $M_{\mathrm{min}} = 10^2 M_{\odot}$  & 0.49 & 4000 \\
High Host Concentration & $1.57$ & $12100$ \\
Low Host Concentration  & $0.91$ & $7300$ \\
\hline
\end{tabular}
\caption{The mean number of signal counts in one year per square
  degree at an angle $\psi=40^\circ$ relative to the Galactic center,
  and the approximate total signal counts on the sky at latitudes
  greater than $b>40^\circ$ for our fiducial model and two models
  which demonstrate the effects of our lack of knowledge of the
  subhalo mass function cutoff scale. The low and high concentration
  models represent extreme models of the host Milky Way properties.}
\label{table:counts}
\end{center}
\end{table}


\Sfig{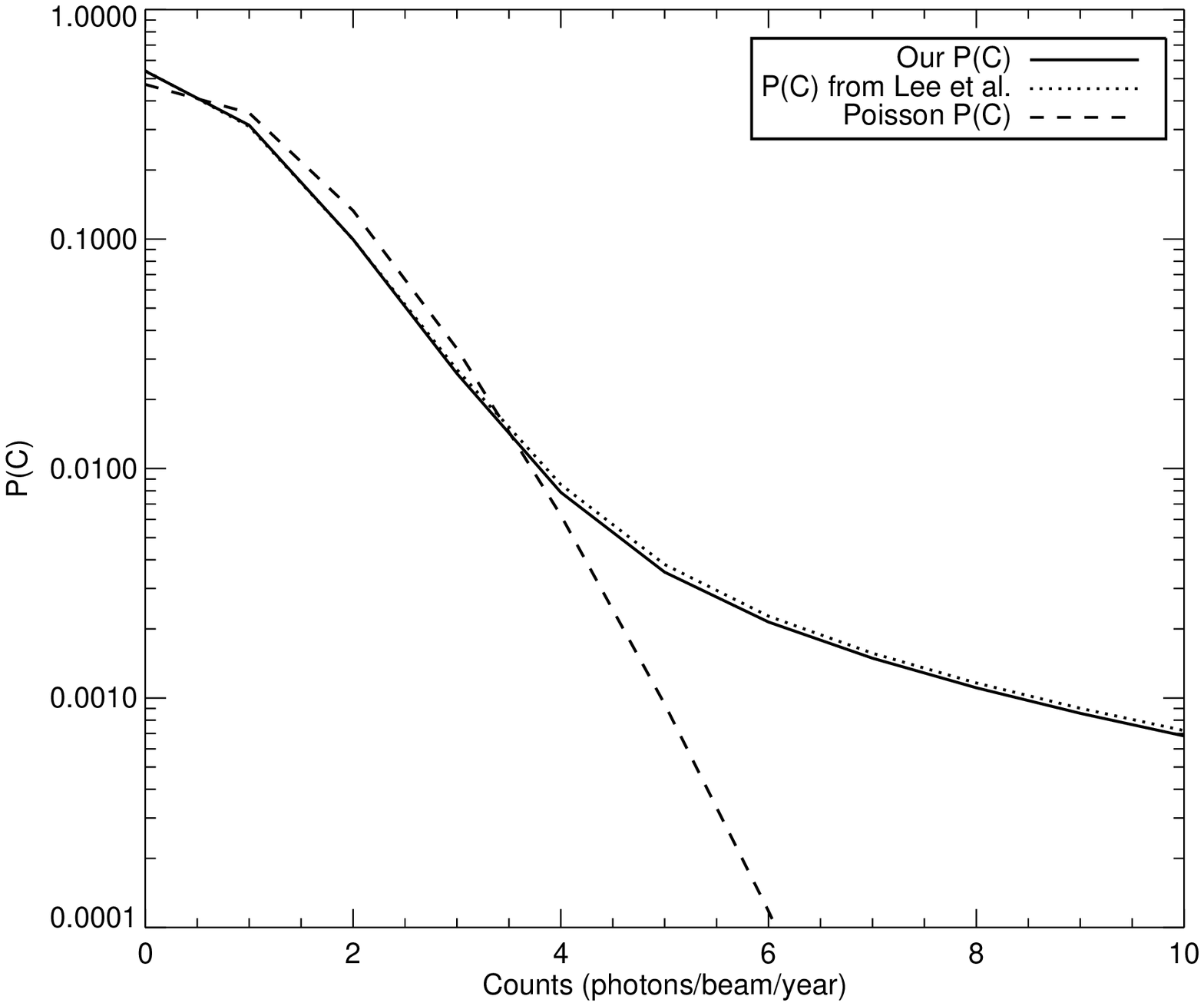}{$P_{sh}(C,\psi_i)$, the probability of observing $C$
  counts in pixel $i$ from all subhalos along the line of sight where
  here $\psi_i=40^\circ$. The $P_{sh}(C,\psi_i)$ predicted by our
  model is compared with that of~\citet{Lee:2008fm}, which we have
  scaled to have the same mean as our model.  The two functions
  are very similar despite the underlying differences of the two
  models.  Both differ significantly from a pure Poisson distribution
  with the same mean number of counts.  }

\Sfig{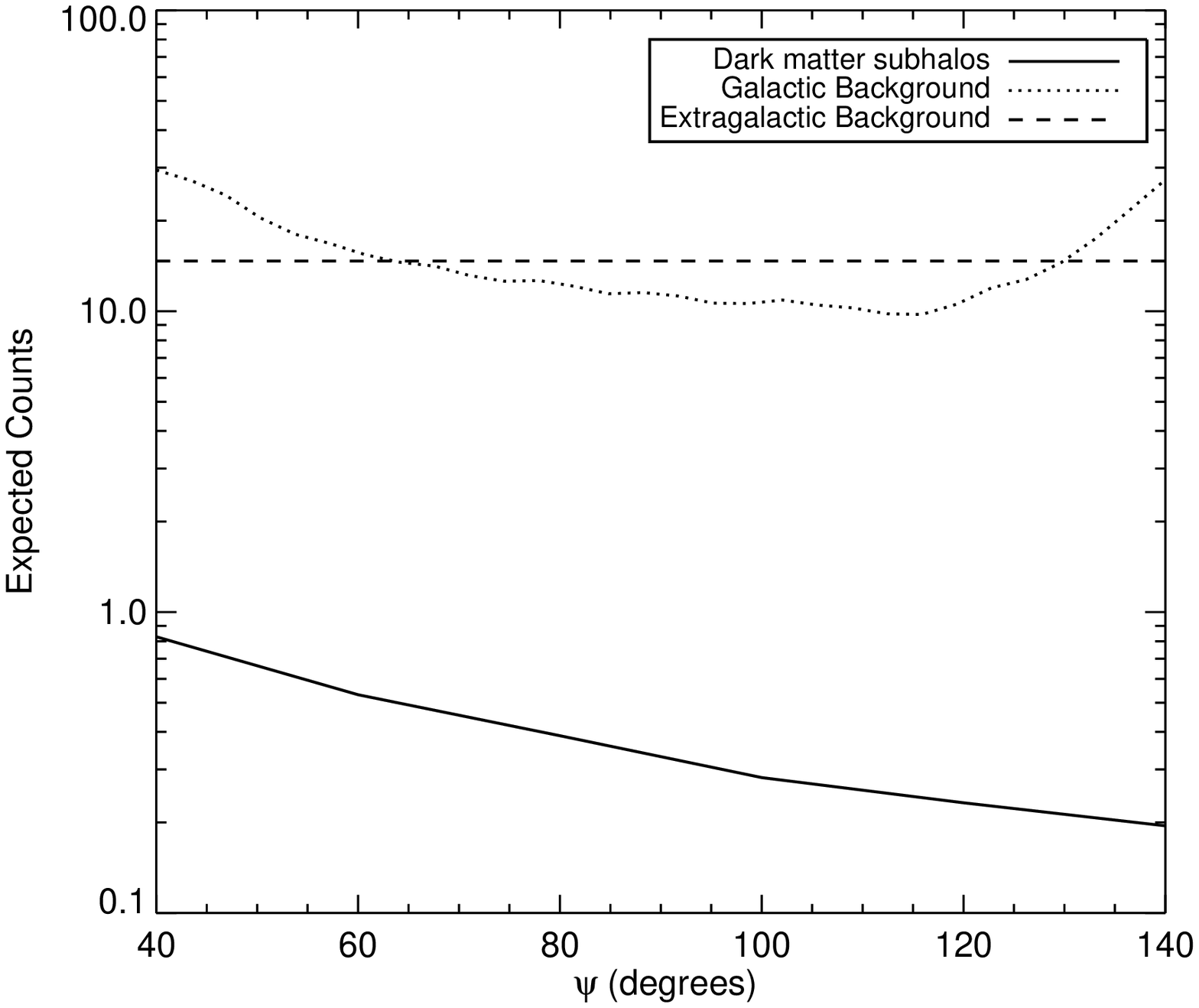}{Expected counts per one square degree pixel above
  1 GeV in 1 year of Fermi data from dark matter annihilation in
  subhalos when $\ppp=10^{-28}$ cm$^3$ s$^{-1}$ GeV$^{-2}$, the
  Galactic background, and the diffuse extra-Galactic background. The
  counts are given in a one square degree pixel.}

\subsection{The Backgrounds: Galactic and Extra-Galactic}
We now move on to discuss the sources of gamma-rays that we consider
in our analysis in addition to the signal from subhalos.  For ease of
book-keeping, we will simply describe gamma-rays from non-subhalo
sources as either Galactic or Extra-Galactic in origin, and now
discuss each of these components in turn.

{\em Galactic Background}-- Cosmic ray interactions with atomic (HI)
and molecular (primarily $H_2$ and CO) gas are the source of diffuse
Galactic gamma-ray emission.  The emission results from the decay of
neutral pions produced in hadronic collisions as well as inverse
Compton scattering of the interstellar radiation field by electrons,
and to a lesser extent bremsstrahlung emission from the interstellar
medium. Accurately modeling this emission is challenging
~\cite{Abdo:2009mr} and indeed crucial for the interpretation and
extraction of a dark matter component in the gamma-ray background.

In our analysis we utilize the standard diffuse gamma-ray emission
model of the LAT science
team~\footnote{http://fermi.gsfc.nasa.gov/ssc/data/access/lat/BackgroundModels.html}.
We take the LAT team model, which is based on the observed
distribution of gas as well as known point sources, as the prediction
of the number of counts in the $i^{th}$ pixel, $C^{\rm b,Fermi}_i$,
for one year of observation. Note that the predicted number of counts
generated by the signal depends only on the angle $\psi_i$ that
separates the pixel from the center of the Galaxy. However the
backgrounds are different: $C^{\rm b,Fermi}_i$ depends on both $l_i$
and $b_i$, and hence not only on $\psi_i$ but also on the azimuthal
position in the annulus.

In each angular pixel, our total number of counts is obtained by
summing over all photons with energy above 1 GeV. We choose this
energy threshold mainly because most of the photons emitted by dark matter pairs
with mass $\sim 100$ GeV or greater are above this energy, and also
because the diffuse Galactic emission is observed to be a steeply
falling power law near these energies. A different choice of energy
threshold is trivial to incorporate into the dark matter model since
it simply corresponds to different $N_\gamma$ in the definition of
$\ppp$, so changing the energy threshold corresponds to changing
$\ppp$.

Using the LAT team diffuse model, we simulate sky maps of diffuse
gamma-ray emission.  When fitting these maps to our model, we
introduce one free parameter, given by the amplitude of the counts
$b_g$. So when considering the diffuse Galactic emission, our model is
simply given by
\begin{equation}
C^{\rm gal}_i = b_g C^{\rm gal,Fermi}_i
\end{equation}
with the true value of $b_g=1$.  

Fig.~\rf{f3} shows the counts from the diffuse Galactic
model in a one square degree pixel as a function of angle from the
Galactic center, $\psi$, with our fiducial normalization
($b_g=1$). Also plotted is the expected signal flux from dark matter
subhalos in equally sized pixels.  For all angles the counts from the
Galactic model are at least an order of magnitude greater than the
counts from subhalos.  As expected, the signal flux falls off with
increasing $\psi$ because the number density of subhalos decreases
with distance from the galactic center (see Eq.~\ref{eq:massfunc}).
The galactic background increases towards $\psi = 0^{\circ}$ and $\psi
= 180^{\circ}$ because most of the diffuse emission is from the
galactic plane.

{\em Extra-Galactic Background}-- The isotropic component of the LAT
team diffuse model is a result of the emission from extra-Galactic and
instrumental sources.  Over the energy range of $\sim 100$ MeV - $100$
GeV, and for $b > 40^\circ$, the isotropic component ascribed to
extragalactic emission is well fit by a power law with index
$2.41$~\cite{Abdo:2010dk}.  The updated diffuse model indicates that
above $1$ GeV, the normalization of the extra-Galactic component is
comparable to that of the dominant
component of Galactic emission that arises from neutral $\pi$
decay. In our analysis, we will simply model the extra-Galactic
component by a number of counts with an amplitude that is allowed to
be free,
\begin{equation}
C^{\rm eg}_i = b_{eg} C^{\rm eg,Fermi}_i. 
\end{equation}
Fig.~\rf{f3} shows our fiducial normalization ($b_{eg}=1$)
is one in which the extra-Galactic flux is about  $30$ times greater than the subhalo
flux, contributing $\sim15$ counts above a GeV in one square degree pixel.

\section{Simulated Maps}\label{sec:map}

Armed with the probability distribution in \ec{prob}, we can generate
simulated maps of the sky for a given experiment specified by its
exposure, $E_i$.  First, though, we construct a simpler map, shown in
Fig.~\rf{f4} to assess ``by eye'' the impact of the
assumed subhalo PDF. Both maps in Fig.~\rf{f4} are
simulations which include the subhalo dark matter signal only. The
central $40^\circ$ is not used so it is zeroed out in all of our maps,
since it will be dominated by Galactic emission. The top panel map in
Fig.~\rf{f4} is drawn from a model with the same
number of expected counts in every pixel as the model introduced in
\S{II}. The counts in each pixel in this map, however, are drawn from
a Poisson distribution. The bottom panel has photons drawn from the
``true'' dark matter PDF.
Fig.~\rf{f2} showed how different
the subhalo PDF is from Poisson, and Fig.~\rf{f4}
illuminates this difference very graphically. There are a number of pixels
with many counts (of order ten) in marked contrast to the
Poisson map which has no high-count pixels.

  \Mapfig{1.0}{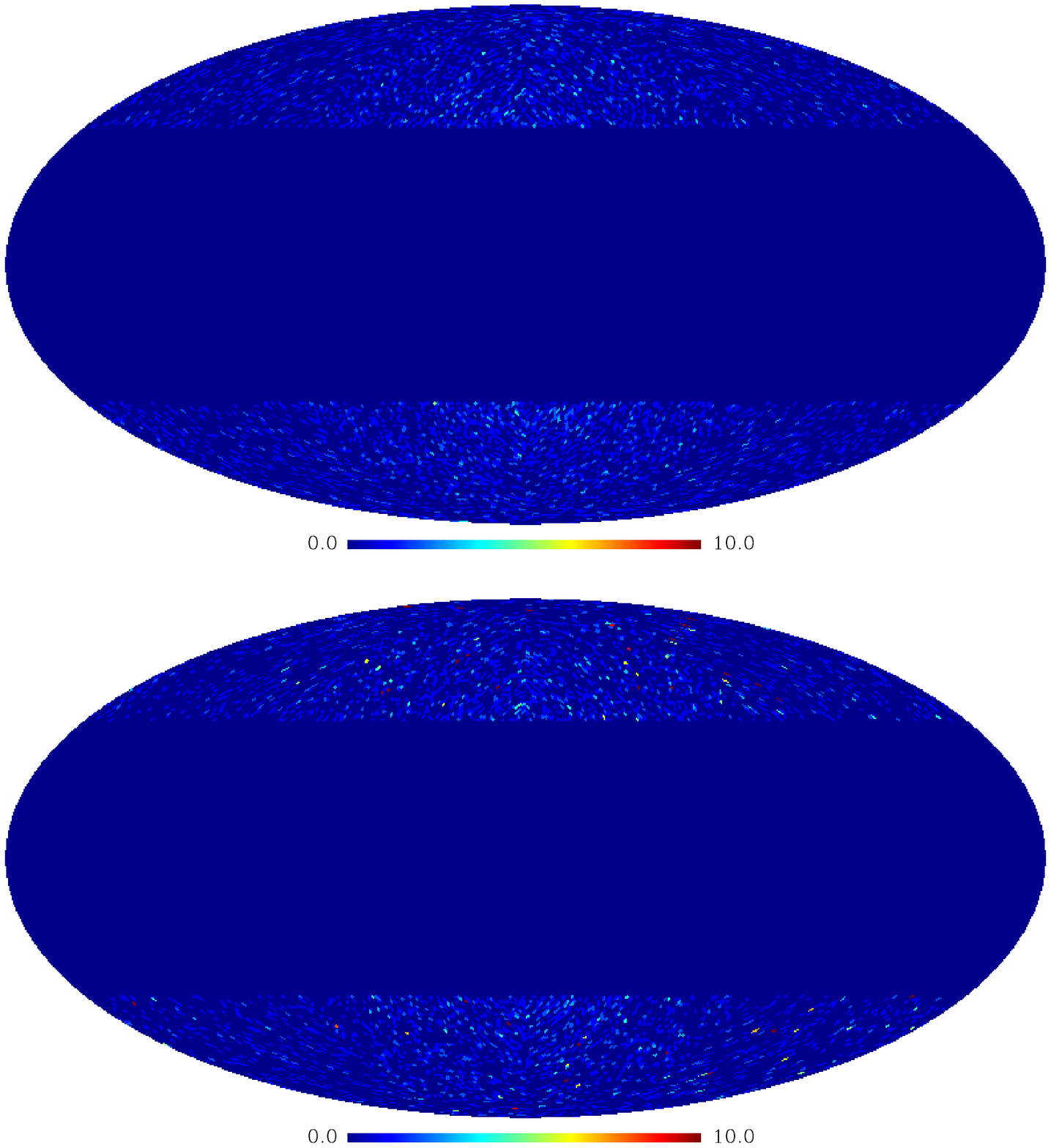}{Simulated maps of the photons above one GeV in
    Fermi produced by the annihilations of dark matter in
    subhalos. {\it Top panel:} Simulated counts drawn from a Poisson
    distribution with the same number of expected events as the dark
    matter PDF. {\it Bottom panel:} Photons drawn from the dark matter
    PDF.}

This visual impression is hidden in a map with
backgrounds. Fig.~\rf{f5} shows two maps with the same
dark matter counts as in Fig.~\rf{f4}, but with counts
from the backgrounds added in. It is no longer possible to
tell the distributions apart by eye, so a more careful statistical
probe is needed. We analyze the maps in the next section to see if
the subhalo signal can be extracted.


\Mapfig{1.0}{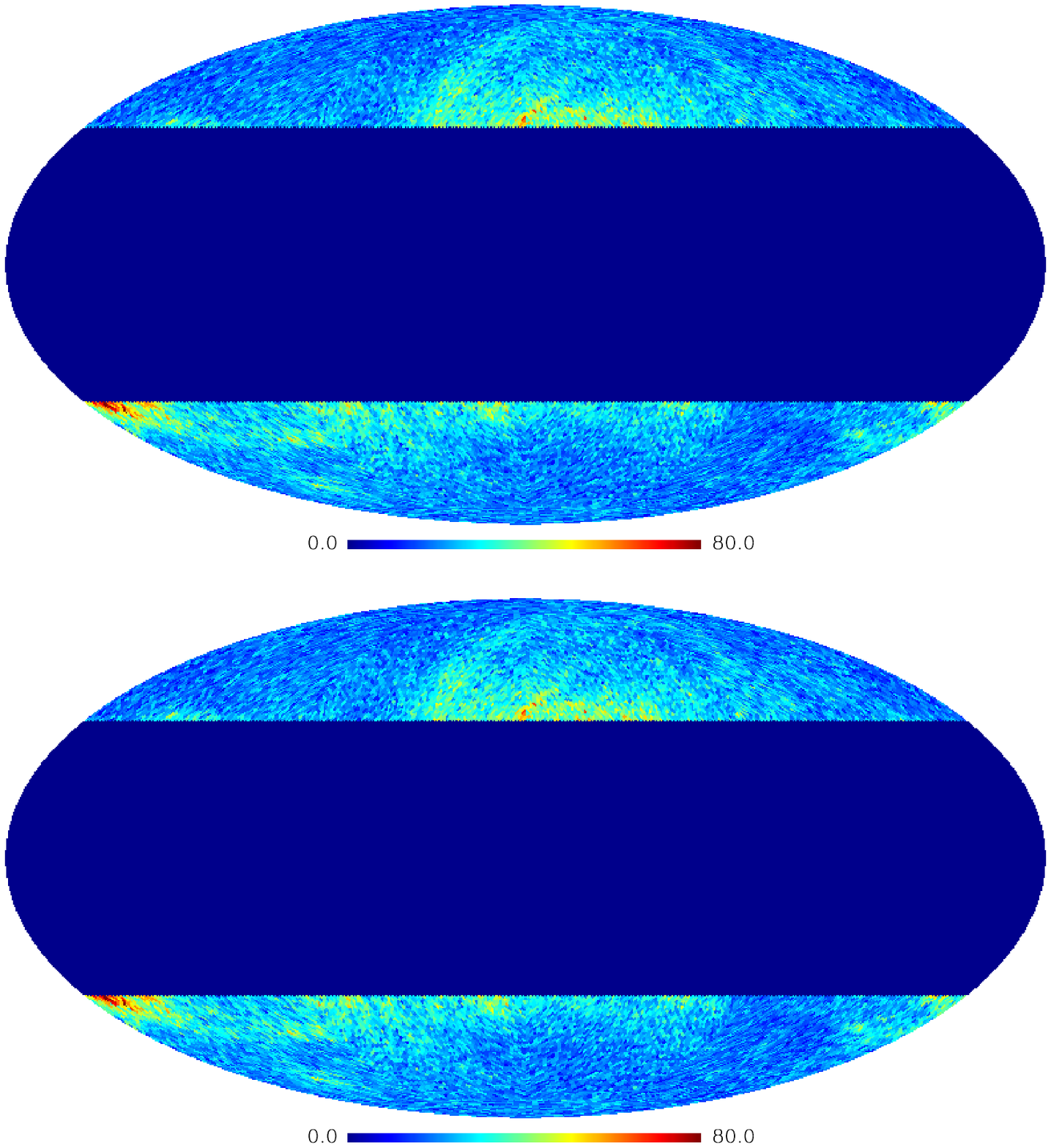}{Same as
  Fig.~\rf{f4} but with the addition of backgrounds
  from the Galaxy and unresolved extra-Galactic sources.}

%

\section{Analysis}\label{sec:chi}

We will analyze the simulated sky constructed in the previous section
in two different ways. First, we will carry out a simple Poisson
analysis to obtain constraints on the parameters. That is, we fit the
data by maximizing a likelihood which assumes (incorrectly) that all
sources of photons are generated from a Poisson distribution, 
\begin{equation}
\mathcal{L}^{\rm Poisson}(\ppp,b_g,b_{eg}) = \prod_{i=1}^{N_p}
\mathcal{P}\left[E_i\bar F_i(\ppp) +C^{\rm g}_i(b_g) +C^{\rm
eg}_i(b_{eg});C_i\right].
\label{eq:Lpoisson} 
\end{equation}
In Eq.~\ref{eq:Lpoisson}, the parameters specifying the amplitudes of
the background are $b_g$ and $b_{eg}$; $C_i$ is the observed number of
counts in pixel $i$ (there are a total of $N_p$ pixels); $\bar F_i$ is
the mean expected flux from dark matter annihilations in pixel $i$;
and $\mathcal{P}[A;B]$ again is the Poisson probability of observing
$B$ counts in a pixel in which the mean expected number of counts is
$A$.
We emphasize that we are (purposely) doing things wrong here: we are
analyzing a map generated from one distribution assuming incorrectly
that the map is Poisson. One of the goals is to determine whether this
flawed (yet simpler) analysis obtains the correct answer. We fit the
data to the three free parameters, find the best fit value in this 3D
space, and then identify 1-, 2-, and 3-sigma constraints by finding
regions within which $\int d\ppp db_g db_{eg} = 0.68,0.95,0.997$. The
best fit value is termed an estimator, the Poisson estimator.

A second way to analyze these simulated maps is to use the ``true''
likelihood. We want to see how much better this approach is than the
Poisson analysis. Here we use the exact likelihood,
\begin{equation}
\mathcal{L} = \prod_{i=1}^{N_p} P(C_i|\ppp,b_g,b_{eg}), 
\end{equation}
where $P(C_i|\ppp,b_g,b_{eg})$ is given in \ec{prob}. Again, we can
form an estimator and allowed regions for the parameters; we call this
estimator the ``true'' or ``exact'' estimator.

We will apply each of these estimators to the signal+background maps
constructed in \S~\ref{sec:map}, but first let us work on the
background-free maps. There it was easy to tell the difference between
the 2 PDF's by eye, so we expect to see considerable differences in
the analyses. Fig.~\rf{f6} show the results of ten
runs applying each estimator. The ``true'' likelihood extracts the
correct value accurately and obtains small error bars. In contrast,
the Poisson likelihood consistently mis-estimates the value of
$\ppp$. Apparently, the Poisson estimator is misled by the
many pixels with few counts so systematically shifts the mean number
of counts lower, thereby leading to an under-estimate of $\ppp$.

\Sfig{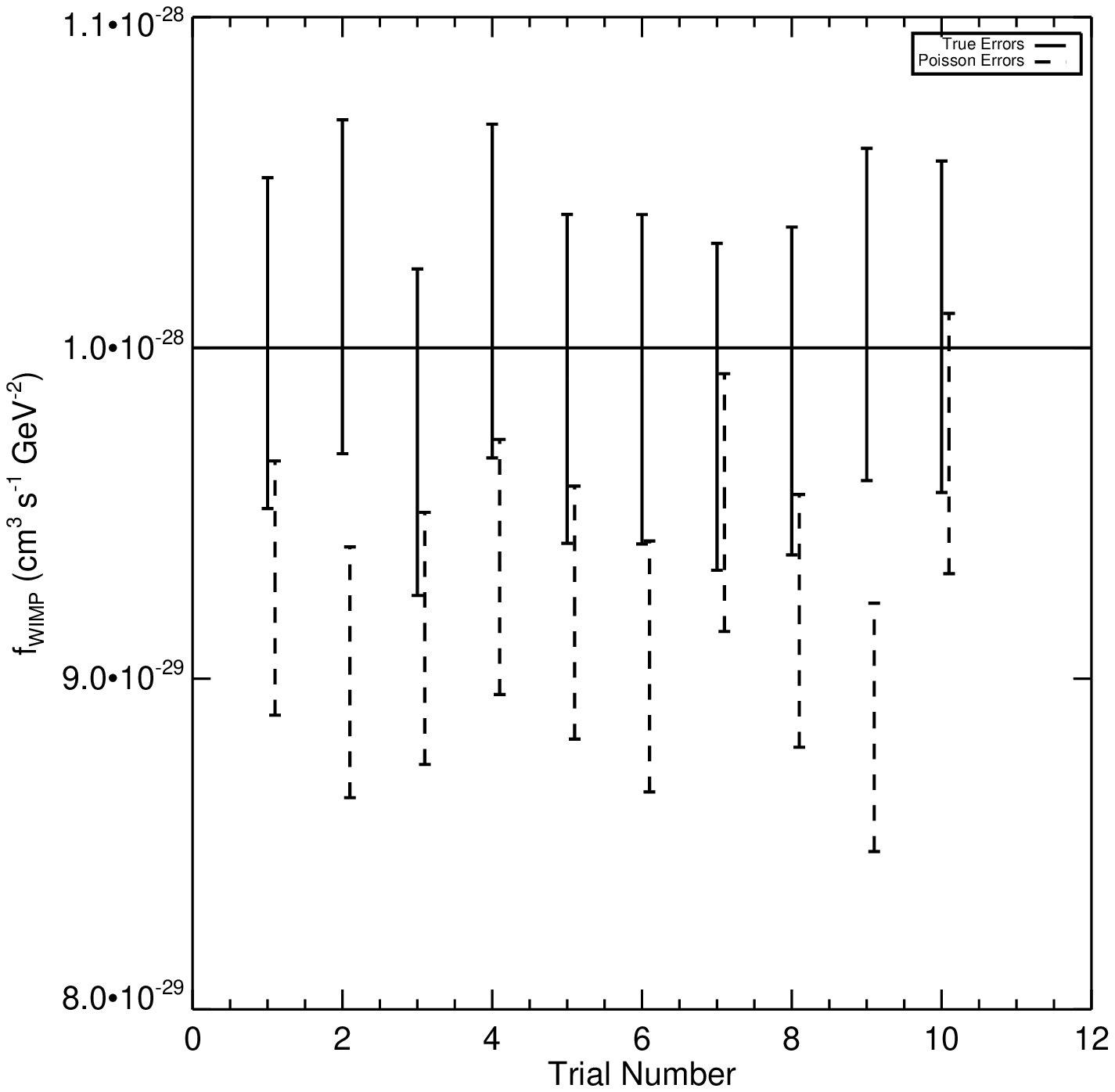}{ Best fit values of $\ppp$ from
  multiple runs when backgrounds are not included. The true likelihood
  recaptures the input value of $\ppp=10^{-28}\,{\rm cm}^3\,{\rm
s}^{-1}\,{\rm GeV}^{-2}$, while the Poisson likelihood systematically
  under-estimates $\ppp$.  The error bars shown represent $3\sigma$
  confidence intervals.}

When backgrounds are added in, it becomes less important to use the
correct PDF. To see this, consider the constraints obtained on one
simulated map using the two estimators, as shown in
Fig.~\rf{f7}. In this realization, both
estimators recapture the true parameter values. The errors on the
parameters are larger when the simpler, Poisson estimator is used, but
the overall impression is that using the Poisson estimator would not
do appreciable damage.  

\Sfigc{.9}{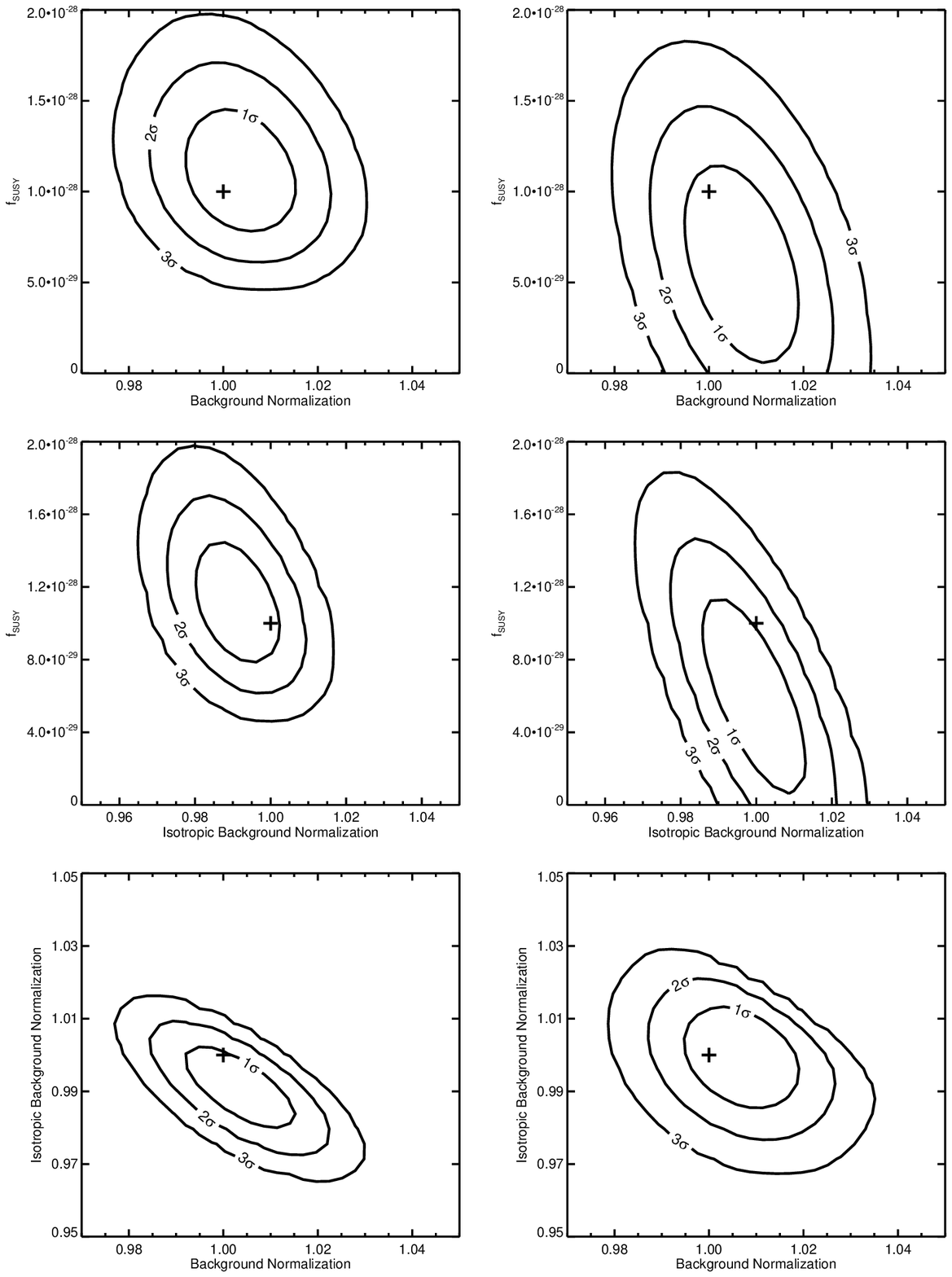}{Constraints from one
  simulated map of signal and backgrounds on the 3 parameters when the
  underlying model has $\ppp=10^{-28}\,{\rm cm}^3\,{\rm s}^{-1}\,{\rm
    GeV}^{-2}$ and $b_g=b_{eg}=1$. {\it Left panels:} Results using
  the ``true'' PDF for dark matter. {\it Right panels:} Constraints
  assuming a Poisson likelihood. The Poisson analysis retrieves the
  correct result even though it assumes the wrong PDF; the allowed
  region is slightly larger if the true likelihood is not known, but
  there is no bias.}

To test this further, we generated 10 such maps.
Fig.~\rf{f8} shows the distribution of best fit values
of $\ppp$, $b_{eg}$, and $b_g$ from these runs. The means are both
close to the true values of the parameters. The errors from the
Poisson analysis are larger by 50 percent, so knowing the PDF does
help, but the danger of a bias appears to be eliminated.

For observation times greater than the one year that we assume in our
analysis, the size of the error contours will of course decrease.
Since $\ppp$ is proportional to the photon flux and since we have
shown that the errors in $\ppp$ are reasonably described by
Poisson statistics, the fractional error in our determination of
$\ppp$ will scale in inverse proportion to the square root of the
exposure time.  Therefore, with data covering the five year expected
lifetime of the Fermi Telescope, we expect the error contours on
$\ppp$ to be about $55\%$ smaller than shown here.

\Sfig{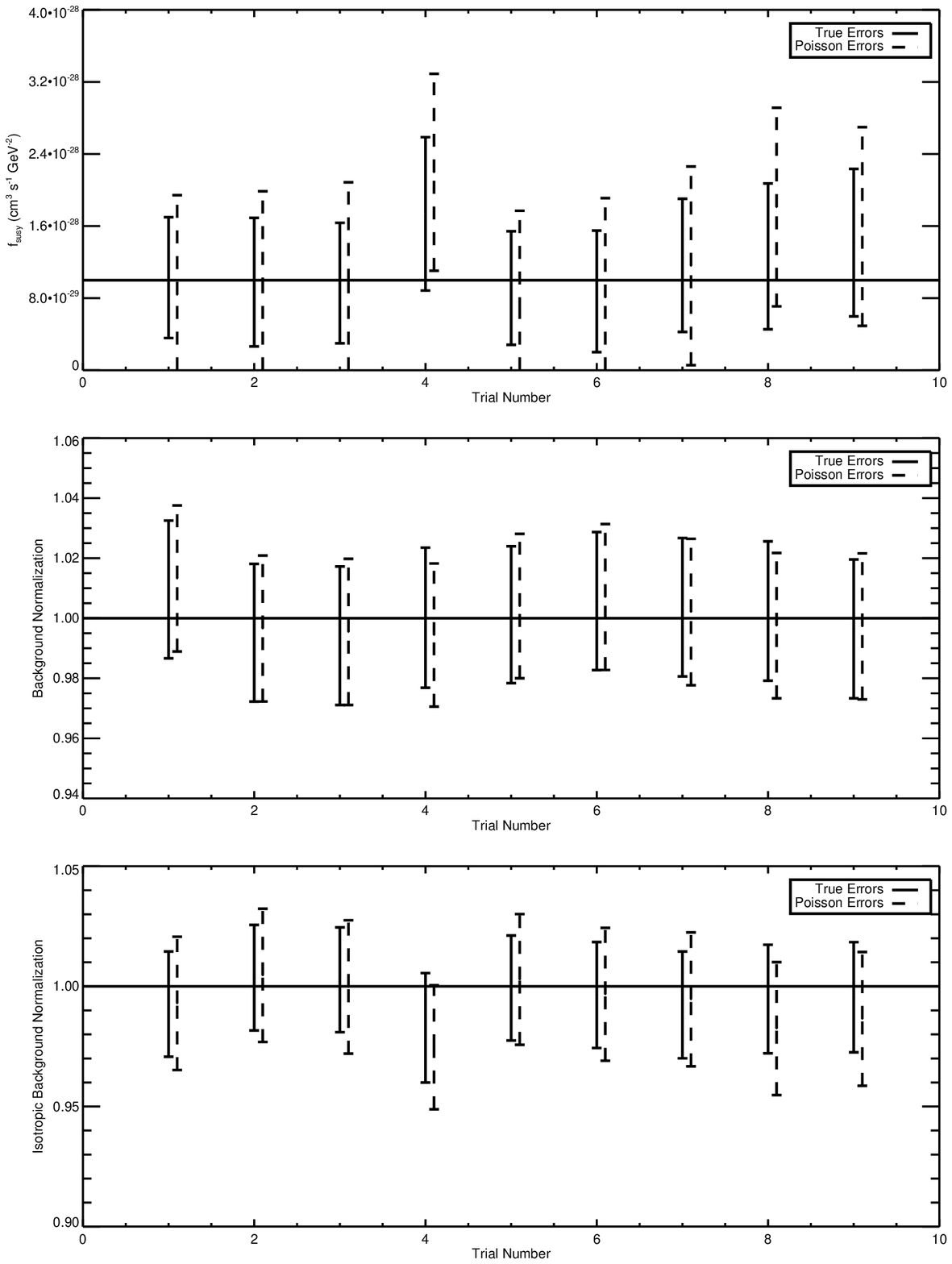}{Best fit values of $\ppp$ and background
  amplitudes from multiple runs. The true values ($\ppp=10^{-28}$
  cm$^3$ s$^{-1}$ GeV$^{-2}$, $b_g=1,b_{eg}=1$) are recaptured by both
  estimators, but the errors -- especially on $\ppp$ -- are about 50
  percent larger when the Poisson estimator is used.  The error bars
  shown represent $3\sigma$ confidence intervals.}


\section{Conclusions}\label{sec:con}

The gamma ray signal from annihilation of Galactic dark matter
subhalos has a probability distribution function which is very
different from a Poisson distribution with the same number of mean
counts. This feature, initially explored in Ref.~\cite{Lee:2008fm} and
fleshed out here with a slightly less restrictive model, should
produce in Fermi many pixels with zero or small number of counts but a
finite set with large number of counts. We have addressed here the
question of how this PDF will affect future analyses and concluded
that, once the backgrounds are added in, a simple analysis which
assumes a Poisson PDF is unbiased and only slightly less powerful than
one which uses the full, correct PDF.

To some extent this is good news: there is a tension between analyses
which are agnostic as to the nature of the signal and those which
assume that many of its underlying features are known and are simply
fitting for parameters. Those in the first class are more robust and
believable because they are based are fewer assumptions; those in the
second are more powerful statistically and will lead to tighter
constraints on the properties of dark matter. When we find little loss
in statistical power from dropping an assumption and moving towards
more agnostic estimators, we should become more optimistic about our
chances of extracting a signal hidden in backgrounds. This is perhaps
the most important result of this work.

Lingering in our discussion, and in the literature at large~(see,
e.g., \cite{Baltz:2008wd}), is the question how much information will
the data contain? The answer to this is encoded in the likelihood
function, and our conclusions are that one year of Fermi data contains
enough information to detect a value of $\ppp=10^{-28}$ cm$^3$
s$^{-1}$ GeV$^{-2}$. We are not claiming that a robust detection of
this small a signal can be expected (for an incomplete sample of
analyses,
see~\cite{Cirelli:2009dv,Abazajian:2010sq,Abdo:2010dk,Hutsi:2010}),
but just that the information is there and we should attempt to
extract it.

Our analysis has included/assumed two types of information about the
signal and backgrounds: the angular distribution and the PDF. We have
not included two other potential discriminants: the spectral shapes of
the different components and the angular two-point functions. The
former is easy to include within the formalism developed here, and we
plan to address this in future work. The latter has been explored by a
number of
authors~\cite{Ando:2005xg,Ando:2006mt,Ando:2006cr,SiegalGaskins:2008ge,SiegalGaskins:2009ux,Hensley:2009gh}
in the form of the $C_l$'s. There is a connection between our work and
the fluctuations explored elsewhere: we have implicitly assumed a flat
$C_l$ spectrum, but one that has a larger amplitude than Poisson
(because the PDF is not Poisson). Whether or not this set of
assumptions includes all of the effects explored elsewhere is an open
question.

\vspace{1cm} 

{\bf Acknowledgments}--
This work has been supported by the National Science Foundation Grant
AST-0908072 and by the US Department of Energy, including grant
DE-FG02-95ER40896.  Support for LS for this work was provided by NASA
through Hubble Fellowship grant HF-51248.01-A awarded by the Space
Telescope Science Institute, which is operated by the Association of
Universities for Research in Astronomy, Inc., for NASA, under contract
NAS 5-26555.


\begin{thebibliography}{51}
\expandafter\ifx\csname natexlab\endcsname\relax\def\natexlab#1{#1}\fi
\expandafter\ifx\csname bibnamefont\endcsname\relax
  \def\bibnamefont#1{#1}\fi
\expandafter\ifx\csname bibfnamefont\endcsname\relax
  \def\bibfnamefont#1{#1}\fi
\expandafter\ifx\csname citenamefont\endcsname\relax
  \def\citenamefont#1{#1}\fi
\expandafter\ifx\csname url\endcsname\relax
  \def\url#1{\texttt{#1}}\fi
\expandafter\ifx\csname urlprefix\endcsname\relax\def\urlprefix{URL }\fi
\providecommand{\bibinfo}[2]{#2}
\providecommand{\eprint}[2][]{\url{#2}}

\bibitem[{\citenamefont{Bertone et~al.}(2005)\citenamefont{Bertone, Hooper, and
  Silk}}]{Bertone:2004pz}
\bibinfo{author}{\bibfnamefont{G.}~\bibnamefont{Bertone}},
  \bibinfo{author}{\bibfnamefont{D.}~\bibnamefont{Hooper}}, \bibnamefont{and}
  \bibinfo{author}{\bibfnamefont{J.}~\bibnamefont{Silk}},
  \bibinfo{journal}{Phys. Rept.} \textbf{\bibinfo{volume}{405}},
  \bibinfo{pages}{279} (\bibinfo{year}{2005}), \eprint{hep-ph/0404175}.

\bibitem[{\citenamefont{{Birkedal} et~al.}(2004)\citenamefont{{Birkedal},
  {Matchev}, and {Perelstein}}}]{Birkedal:2004}
\bibinfo{author}{\bibfnamefont{A.}~\bibnamefont{{Birkedal}}},
  \bibinfo{author}{\bibfnamefont{K.}~\bibnamefont{{Matchev}}},
  \bibnamefont{and}
  \bibinfo{author}{\bibfnamefont{M.}~\bibnamefont{{Perelstein}}},
  \bibinfo{journal}{\prd} \textbf{\bibinfo{volume}{70}},
  \bibinfo{pages}{077701} (\bibinfo{year}{2004}),
  \eprint{arXiv:hep-ph/0403004}.

\bibitem[{\citenamefont{{Feng} et~al.}(2006)\citenamefont{{Feng}, {Su}, and
  {Takayama}}}]{Feng:2006}
\bibinfo{author}{\bibfnamefont{J.~L.} \bibnamefont{{Feng}}},
  \bibinfo{author}{\bibfnamefont{S.}~\bibnamefont{{Su}}}, \bibnamefont{and}
  \bibinfo{author}{\bibfnamefont{F.}~\bibnamefont{{Takayama}}},
  \bibinfo{journal}{Physical Review Letters} \textbf{\bibinfo{volume}{96}},
  \bibinfo{pages}{151802} (\bibinfo{year}{2006}),
  \eprint{arXiv:hep-ph/0503117}.

\bibitem[{\citenamefont{{Drukier} and {Stodolsky}}(1984)}]{Drukier:1984}
\bibinfo{author}{\bibfnamefont{A.}~\bibnamefont{{Drukier}}} \bibnamefont{and}
  \bibinfo{author}{\bibfnamefont{L.}~\bibnamefont{{Stodolsky}}},
  \bibinfo{journal}{\prd} \textbf{\bibinfo{volume}{30}}, \bibinfo{pages}{2295}
  (\bibinfo{year}{1984}).

\bibitem[{\citenamefont{{Goodman} and {Witten}}(1985)}]{Goodman:1985}
\bibinfo{author}{\bibfnamefont{M.~W.} \bibnamefont{{Goodman}}}
  \bibnamefont{and} \bibinfo{author}{\bibfnamefont{E.}~\bibnamefont{{Witten}}},
  \bibinfo{journal}{\prd} \textbf{\bibinfo{volume}{31}}, \bibinfo{pages}{3059}
  (\bibinfo{year}{1985}).

\bibitem[{\citenamefont{{Silk} and {Srednicki}}(1984)}]{Silk:1984}
\bibinfo{author}{\bibfnamefont{J.}~\bibnamefont{{Silk}}} \bibnamefont{and}
  \bibinfo{author}{\bibfnamefont{M.}~\bibnamefont{{Srednicki}}},
  \bibinfo{journal}{Physical Review Letters} \textbf{\bibinfo{volume}{53}},
  \bibinfo{pages}{624} (\bibinfo{year}{1984}).

\bibitem[{\citenamefont{{Gunn} et~al.}(1978)\citenamefont{{Gunn}, {Lee},
  {Lerche}, {Schramm}, and {Steigman}}}]{Gunn:1978}
\bibinfo{author}{\bibfnamefont{J.~E.} \bibnamefont{{Gunn}}},
  \bibinfo{author}{\bibfnamefont{B.~W.} \bibnamefont{{Lee}}},
  \bibinfo{author}{\bibfnamefont{I.}~\bibnamefont{{Lerche}}},
  \bibinfo{author}{\bibfnamefont{D.~N.} \bibnamefont{{Schramm}}},
  \bibnamefont{and}
  \bibinfo{author}{\bibfnamefont{G.}~\bibnamefont{{Steigman}}},
  \bibinfo{journal}{\apj} \textbf{\bibinfo{volume}{223}}, \bibinfo{pages}{1015}
  (\bibinfo{year}{1978}).

\bibitem[{\citenamefont{{Golubkov} et~al.}(1999)\citenamefont{{Golubkov},
  {Konoplich}, {Mignani}, {Fargion}, and {Khlopov}}}]{Golubkov:1999}
\bibinfo{author}{\bibfnamefont{Y.~A.} \bibnamefont{{Golubkov}}},
  \bibinfo{author}{\bibfnamefont{R.~V.} \bibnamefont{{Konoplich}}},
  \bibinfo{author}{\bibfnamefont{R.}~\bibnamefont{{Mignani}}},
  \bibinfo{author}{\bibfnamefont{D.}~\bibnamefont{{Fargion}}},
  \bibnamefont{and} \bibinfo{author}{\bibfnamefont{M.~Y.}
  \bibnamefont{{Khlopov}}}, \bibinfo{journal}{Soviet Journal of Experimental
  and Theoretical Physics Letters} \textbf{\bibinfo{volume}{69}},
  \bibinfo{pages}{434} (\bibinfo{year}{1999}), \eprint{arXiv:astro-ph/9903086}.

\bibitem[{\citenamefont{Baltz et~al.}(2008)}]{Baltz:2008wd}
\bibinfo{author}{\bibfnamefont{E.~A.} \bibnamefont{Baltz}}
  \bibnamefont{et~al.}, \bibinfo{journal}{JCAP}
  \textbf{\bibinfo{volume}{0807}}, \bibinfo{pages}{013} (\bibinfo{year}{2008}),
  \eprint{0806.2911}.

\bibitem[{\citenamefont{Hui and Collaboration}(2009)}]{Hui:2008tt}
\bibinfo{author}{\bibfnamefont{C.~M.} \bibnamefont{Hui}} \bibnamefont{and}
  \bibinfo{author}{\bibfnamefont{f.~t.~V.} \bibnamefont{Collaboration}},
  \bibinfo{journal}{AIP Conf. Proc.} \textbf{\bibinfo{volume}{1085}},
  \bibinfo{pages}{407} (\bibinfo{year}{2009}), \eprint{0810.1913}.

\bibitem[{\citenamefont{Adriani et~al.}(2009)}]{Adriani:2008zr}
\bibinfo{author}{\bibfnamefont{O.}~\bibnamefont{Adriani}} \bibnamefont{et~al.}
  (\bibinfo{collaboration}{PAMELA}), \bibinfo{journal}{Nature}
  \textbf{\bibinfo{volume}{458}}, \bibinfo{pages}{607} (\bibinfo{year}{2009}),
  \eprint{0810.4995}.

\bibitem[{\citenamefont{Resconi and Collaboration}(2009)}]{Resconi:2008fe}
\bibinfo{author}{\bibfnamefont{E.}~\bibnamefont{Resconi}} \bibnamefont{and}
  \bibinfo{author}{\bibfnamefont{f.~t.~I.} \bibnamefont{Collaboration}},
  \bibinfo{journal}{Nucl. Instrum. Meth.} \textbf{\bibinfo{volume}{A602}},
  \bibinfo{pages}{7} (\bibinfo{year}{2009}), \eprint{0807.3891}.

\bibitem[{\citenamefont{Bergstrom et~al.}(1998)\citenamefont{Bergstrom, Ullio,
  and Buckley}}]{Bergstrom:1997fj}
\bibinfo{author}{\bibfnamefont{L.}~\bibnamefont{Bergstrom}},
  \bibinfo{author}{\bibfnamefont{P.}~\bibnamefont{Ullio}}, \bibnamefont{and}
  \bibinfo{author}{\bibfnamefont{J.~H.} \bibnamefont{Buckley}},
  \bibinfo{journal}{Astropart. Phys.} \textbf{\bibinfo{volume}{9}},
  \bibinfo{pages}{137} (\bibinfo{year}{1998}), \eprint{astro-ph/9712318}.

\bibitem[{\citenamefont{Fornengo et~al.}(2004)\citenamefont{Fornengo, Pieri,
  and Scopel}}]{Fornengo:2004kj}
\bibinfo{author}{\bibfnamefont{N.}~\bibnamefont{Fornengo}},
  \bibinfo{author}{\bibfnamefont{L.}~\bibnamefont{Pieri}}, \bibnamefont{and}
  \bibinfo{author}{\bibfnamefont{S.}~\bibnamefont{Scopel}},
  \bibinfo{journal}{Phys. Rev.} \textbf{\bibinfo{volume}{D70}},
  \bibinfo{pages}{103529} (\bibinfo{year}{2004}), \eprint{hep-ph/0407342}.

\bibitem[{\citenamefont{Baltz et~al.}(2007)\citenamefont{Baltz, Taylor, and
  Wai}}]{Baltz:2006sv}
\bibinfo{author}{\bibfnamefont{E.~A.} \bibnamefont{Baltz}},
  \bibinfo{author}{\bibfnamefont{J.~E.} \bibnamefont{Taylor}},
  \bibnamefont{and} \bibinfo{author}{\bibfnamefont{L.~L.} \bibnamefont{Wai}},
  \bibinfo{journal}{Astrophys. J. Lett.} \textbf{\bibinfo{volume}{659}},
  \bibinfo{pages}{L125} (\bibinfo{year}{2007}), \eprint{astro-ph/0610731}.

\bibitem[{\citenamefont{Venters and Pavlidou}(2007)}]{Venters:2007rn}
\bibinfo{author}{\bibfnamefont{T.~M.} \bibnamefont{Venters}} \bibnamefont{and}
  \bibinfo{author}{\bibfnamefont{V.}~\bibnamefont{Pavlidou}},
  \bibinfo{journal}{AIP Conf. Proc.} \textbf{\bibinfo{volume}{921}},
  \bibinfo{pages}{163} (\bibinfo{year}{2007}), \eprint{0704.2417}.

\bibitem[{\citenamefont{Hooper and Serpico}(2007)}]{Hooper:2007be}
\bibinfo{author}{\bibfnamefont{D.}~\bibnamefont{Hooper}} \bibnamefont{and}
  \bibinfo{author}{\bibfnamefont{P.~D.} \bibnamefont{Serpico}},
  \bibinfo{journal}{JCAP} \textbf{\bibinfo{volume}{0706}}, \bibinfo{pages}{013}
  (\bibinfo{year}{2007}), \eprint{astro-ph/0702328}.

\bibitem[{\citenamefont{{Cuoco} et~al.}(2008)\citenamefont{{Cuoco},
  {Brandbyge}, {Hannestad}, {Haugb{\o}lle}, and {Miele}}}]{Cuoco:2007sh}
\bibinfo{author}{\bibfnamefont{A.}~\bibnamefont{{Cuoco}}},
  \bibinfo{author}{\bibfnamefont{J.}~\bibnamefont{{Brandbyge}}},
  \bibinfo{author}{\bibfnamefont{S.}~\bibnamefont{{Hannestad}}},
  \bibinfo{author}{\bibfnamefont{T.}~\bibnamefont{{Haugb{\o}lle}}},
  \bibnamefont{and} \bibinfo{author}{\bibfnamefont{G.}~\bibnamefont{{Miele}}},
  \bibinfo{journal}{\prd} \textbf{\bibinfo{volume}{77}},
  \bibinfo{pages}{123518} (\bibinfo{year}{2008}), \eprint{0710.4136}.

\bibitem[{\citenamefont{Dodelson et~al.}(2008)\citenamefont{Dodelson, Hooper,
  and Serpico}}]{Dodelson:2007gd}
\bibinfo{author}{\bibfnamefont{S.}~\bibnamefont{Dodelson}},
  \bibinfo{author}{\bibfnamefont{D.}~\bibnamefont{Hooper}}, \bibnamefont{and}
  \bibinfo{author}{\bibfnamefont{P.~D.} \bibnamefont{Serpico}},
  \bibinfo{journal}{Phys. Rev.} \textbf{\bibinfo{volume}{D77}},
  \bibinfo{pages}{063512} (\bibinfo{year}{2008}), \eprint{0711.4621}.

\bibitem[{\citenamefont{Kuhlen et~al.}(2007)\citenamefont{Kuhlen, Diemand, and
  Madau}}]{Kuhlen:2007wv}
\bibinfo{author}{\bibfnamefont{M.}~\bibnamefont{Kuhlen}},
  \bibinfo{author}{\bibfnamefont{J.}~\bibnamefont{Diemand}}, \bibnamefont{and}
  \bibinfo{author}{\bibfnamefont{P.}~\bibnamefont{Madau}},
  \bibinfo{journal}{AIP Conf. Proc.} \textbf{\bibinfo{volume}{921}},
  \bibinfo{pages}{135} (\bibinfo{year}{2007}), \eprint{0704.0944}.

\bibitem[{\citenamefont{Kuhlen et~al.}(2008)\citenamefont{Kuhlen, Diemand, and
  Madau}}]{Kuhlen:2008aw}
\bibinfo{author}{\bibfnamefont{M.}~\bibnamefont{Kuhlen}},
  \bibinfo{author}{\bibfnamefont{J.}~\bibnamefont{Diemand}}, \bibnamefont{and}
  \bibinfo{author}{\bibfnamefont{P.}~\bibnamefont{Madau}}
  (\bibinfo{year}{2008}), \eprint{0805.4416}.

\bibitem[{\citenamefont{Stecker and Salamon}(2001)}]{Stecker:2001dk}
\bibinfo{author}{\bibfnamefont{F.~W.} \bibnamefont{Stecker}} \bibnamefont{and}
  \bibinfo{author}{\bibfnamefont{M.~H.} \bibnamefont{Salamon}}
  (\bibinfo{year}{2001}), \eprint{astro-ph/0104368}.

\bibitem[{\citenamefont{Dermer}(2007)}]{Dermer:2007fg}
\bibinfo{author}{\bibfnamefont{C.~D.} \bibnamefont{Dermer}},
  \bibinfo{journal}{AIP Conf. Proc.} \textbf{\bibinfo{volume}{921}},
  \bibinfo{pages}{122} (\bibinfo{year}{2007}), \eprint{0704.2888}.

\bibitem[{\citenamefont{Lee et~al.}(2008)\citenamefont{Lee, Ando, and
  Kamionkowski}}]{Lee:2008fm}
\bibinfo{author}{\bibfnamefont{S.~K.} \bibnamefont{Lee}},
  \bibinfo{author}{\bibfnamefont{S.}~\bibnamefont{Ando}}, \bibnamefont{and}
  \bibinfo{author}{\bibfnamefont{M.}~\bibnamefont{Kamionkowski}}
  (\bibinfo{year}{2008}), \eprint{0810.1284}.

\bibitem[{\citenamefont{Dodelson et~al.}(2009)\citenamefont{Dodelson, Belikov,
  Hooper, and Serpico}}]{Dodelson:2009ih}
\bibinfo{author}{\bibfnamefont{S.}~\bibnamefont{Dodelson}},
  \bibinfo{author}{\bibfnamefont{A.~V.} \bibnamefont{Belikov}},
  \bibinfo{author}{\bibfnamefont{D.}~\bibnamefont{Hooper}}, \bibnamefont{and}
  \bibinfo{author}{\bibfnamefont{P.}~\bibnamefont{Serpico}},
  \bibinfo{journal}{Phys. Rev.} \textbf{\bibinfo{volume}{D80}},
  \bibinfo{pages}{083504} (\bibinfo{year}{2009}), \eprint{0903.2829}.

\bibitem[{\citenamefont{Siegal-Gaskins and
  Pavlidou}(2009)}]{SiegalGaskins:2009ux}
\bibinfo{author}{\bibfnamefont{J.~M.} \bibnamefont{Siegal-Gaskins}}
  \bibnamefont{and} \bibinfo{author}{\bibfnamefont{V.}~\bibnamefont{Pavlidou}}
  (\bibinfo{year}{2009}), \eprint{0901.3776}.

\bibitem[{\citenamefont{Springel et~al.}(2008{\natexlab{a}})}]{Springel:2008zz}
\bibinfo{author}{\bibfnamefont{V.}~\bibnamefont{Springel}}
  \bibnamefont{et~al.}, \bibinfo{journal}{Nature}
  \textbf{\bibinfo{volume}{456N7218}}, \bibinfo{pages}{73}
  (\bibinfo{year}{2008}{\natexlab{a}}).

\bibitem[{\citenamefont{{Hofmann} et~al.}(2001)\citenamefont{{Hofmann},
  {Schwarz}, and {St{\"o}cker}}}]{HSS01}
\bibinfo{author}{\bibfnamefont{S.}~\bibnamefont{{Hofmann}}},
  \bibinfo{author}{\bibfnamefont{D.~J.} \bibnamefont{{Schwarz}}},
  \bibnamefont{and}
  \bibinfo{author}{\bibfnamefont{H.}~\bibnamefont{{St{\"o}cker}}},
  \bibinfo{journal}{\prd} \textbf{\bibinfo{volume}{64}},
  \bibinfo{pages}{083507} (\bibinfo{year}{2001}),
  \eprint{arXiv:astro-ph/0104173}.

\bibitem[{\citenamefont{Chen et~al.}(2001)\citenamefont{Chen, Kamionkowski, and
  Zhang}}]{CKZ02}
\bibinfo{author}{\bibfnamefont{X.}~\bibnamefont{Chen}},
  \bibinfo{author}{\bibfnamefont{M.}~\bibnamefont{Kamionkowski}},
  \bibnamefont{and} \bibinfo{author}{\bibfnamefont{X.}~\bibnamefont{Zhang}},
  \bibinfo{journal}{Phys. Rev.} \textbf{\bibinfo{volume}{D64}},
  \bibinfo{pages}{021302} (\bibinfo{year}{2001}).

\bibitem[{\citenamefont{Profumo et~al.}(2006)\citenamefont{Profumo, Sigurdson,
  and Kamionkowski}}]{Profumo:2006bv}
\bibinfo{author}{\bibfnamefont{S.}~\bibnamefont{Profumo}},
  \bibinfo{author}{\bibfnamefont{K.}~\bibnamefont{Sigurdson}},
  \bibnamefont{and}
  \bibinfo{author}{\bibfnamefont{M.}~\bibnamefont{Kamionkowski}},
  \bibinfo{journal}{Phys. Rev. Lett.} \textbf{\bibinfo{volume}{97}},
  \bibinfo{pages}{031301} (\bibinfo{year}{2006}), \eprint{astro-ph/0603373}.

\bibitem[{\citenamefont{{Schmid} et~al.}(1999)\citenamefont{{Schmid},
  {Schwarz}, and {Widerin}}}]{Schmid:1998mx}
\bibinfo{author}{\bibfnamefont{C.}~\bibnamefont{{Schmid}}},
  \bibinfo{author}{\bibfnamefont{D.~J.} \bibnamefont{{Schwarz}}},
  \bibnamefont{and}
  \bibinfo{author}{\bibfnamefont{P.}~\bibnamefont{{Widerin}}},
  \bibinfo{journal}{\prd} \textbf{\bibinfo{volume}{59}},
  \bibinfo{pages}{043517} (\bibinfo{year}{1999}),
  \eprint{arXiv:astro-ph/9807257}.

\bibitem[{\citenamefont{Green et~al.}(2004)\citenamefont{Green, Hofmann, and
  Schwarz}}]{Green:2003un}
\bibinfo{author}{\bibfnamefont{A.~M.} \bibnamefont{Green}},
  \bibinfo{author}{\bibfnamefont{S.}~\bibnamefont{Hofmann}}, \bibnamefont{and}
  \bibinfo{author}{\bibfnamefont{D.~J.} \bibnamefont{Schwarz}},
  \bibinfo{journal}{Mon. Not. Roy. Astron. Soc.}
  \textbf{\bibinfo{volume}{353}}, \bibinfo{pages}{L23} (\bibinfo{year}{2004}),
  \eprint{astro-ph/0309621}.

\bibitem[{\citenamefont{Green et~al.}(2005)}]{Green:2005fa}
\bibinfo{author}{\bibfnamefont{A.~M.} \bibnamefont{Green}}
  \bibnamefont{et~al.}, \bibinfo{journal}{JCAP}
  \textbf{\bibinfo{volume}{0508}}, \bibinfo{pages}{003} (\bibinfo{year}{2005}),
  \eprint{astro-ph/0503387}.

\bibitem[{\citenamefont{Loeb and Zaldarriaga}(2005)}]{LZ05}
\bibinfo{author}{\bibfnamefont{A.}~\bibnamefont{Loeb}} \bibnamefont{and}
  \bibinfo{author}{\bibfnamefont{M.}~\bibnamefont{Zaldarriaga}},
  \bibinfo{journal}{Phys. Rev.} \textbf{\bibinfo{volume}{D71}},
  \bibinfo{pages}{103520} (\bibinfo{year}{2005}), \eprint{astro-ph/0504112}.

\bibitem[{\citenamefont{Martinez et~al.}(2009)\citenamefont{Martinez, Bullock,
  Kaplinghat, Strigari, and Trotta}}]{Martinez:2009jh}
\bibinfo{author}{\bibfnamefont{G.~D.} \bibnamefont{Martinez}},
  \bibinfo{author}{\bibfnamefont{J.~S.} \bibnamefont{Bullock}},
  \bibinfo{author}{\bibfnamefont{M.}~\bibnamefont{Kaplinghat}},
  \bibinfo{author}{\bibfnamefont{L.~E.} \bibnamefont{Strigari}},
  \bibnamefont{and} \bibinfo{author}{\bibfnamefont{R.}~\bibnamefont{Trotta}},
  \bibinfo{journal}{JCAP} \textbf{\bibinfo{volume}{0906}}, \bibinfo{pages}{014}
  (\bibinfo{year}{2009}), \eprint{0902.4715}.

\bibitem[{\citenamefont{{Koushiappas} et~al.}(2010)\citenamefont{{Koushiappas},
  {Zentner}, and {Kravtsov}}}]{Koushiappas:2010}
\bibinfo{author}{\bibfnamefont{S.~M.} \bibnamefont{{Koushiappas}}},
  \bibinfo{author}{\bibfnamefont{A.~R.} \bibnamefont{{Zentner}}},
  \bibnamefont{and} \bibinfo{author}{\bibfnamefont{A.~V.}
  \bibnamefont{{Kravtsov}}}, \bibinfo{journal}{ArXiv e-prints}
  (\bibinfo{year}{2010}), \eprint{1006.2391}.

\bibitem[{\citenamefont{Bergstrom et~al.}(2001)\citenamefont{Bergstrom, Edsjo,
  and Ullio}}]{Bergstrom:2001jj}
\bibinfo{author}{\bibfnamefont{L.}~\bibnamefont{Bergstrom}},
  \bibinfo{author}{\bibfnamefont{J.}~\bibnamefont{Edsjo}}, \bibnamefont{and}
  \bibinfo{author}{\bibfnamefont{P.}~\bibnamefont{Ullio}},
  \bibinfo{journal}{Phys. Rev. Lett.} \textbf{\bibinfo{volume}{87}},
  \bibinfo{pages}{251301} (\bibinfo{year}{2001}), \eprint{astro-ph/0105048}.

\bibitem[{\citenamefont{Strigari et~al.}(2007)\citenamefont{Strigari,
  Koushiappas, Bullock, and Kaplinghat}}]{Strigari:2006rd}
\bibinfo{author}{\bibfnamefont{L.~E.} \bibnamefont{Strigari}},
  \bibinfo{author}{\bibfnamefont{S.~M.} \bibnamefont{Koushiappas}},
  \bibinfo{author}{\bibfnamefont{J.~S.} \bibnamefont{Bullock}},
  \bibnamefont{and}
  \bibinfo{author}{\bibfnamefont{M.}~\bibnamefont{Kaplinghat}},
  \bibinfo{journal}{Phys. Rev.} \textbf{\bibinfo{volume}{D75}},
  \bibinfo{pages}{083526} (\bibinfo{year}{2007}), \eprint{astro-ph/0611925}.

\bibitem[{\citenamefont{Diemand et~al.}(2007)\citenamefont{Diemand, Kuhlen, and
  Madau}}]{Diemand:2006ik}
\bibinfo{author}{\bibfnamefont{J.}~\bibnamefont{Diemand}},
  \bibinfo{author}{\bibfnamefont{M.}~\bibnamefont{Kuhlen}}, \bibnamefont{and}
  \bibinfo{author}{\bibfnamefont{P.}~\bibnamefont{Madau}},
  \bibinfo{journal}{Astrophys. J.} \textbf{\bibinfo{volume}{657}},
  \bibinfo{pages}{262} (\bibinfo{year}{2007}), \eprint{astro-ph/0611370}.

\bibitem[{\citenamefont{Springel et~al.}(2008{\natexlab{b}})}]{Springel:2008cc}
\bibinfo{author}{\bibfnamefont{V.}~\bibnamefont{Springel}}
  \bibnamefont{et~al.}, \bibinfo{journal}{Mon. Not. Roy. Astron. Soc.}
  \textbf{\bibinfo{volume}{391}}, \bibinfo{pages}{1685}
  (\bibinfo{year}{2008}{\natexlab{b}}), \eprint{0809.0898}.

\bibitem[{\citenamefont{D'Onghia et~al.}(2010)\citenamefont{D'Onghia, Springel,
  Hernquist, and Keres}}]{D'Onghia:2009pz}
\bibinfo{author}{\bibfnamefont{E.}~\bibnamefont{D'Onghia}},
  \bibinfo{author}{\bibfnamefont{V.}~\bibnamefont{Springel}},
  \bibinfo{author}{\bibfnamefont{L.}~\bibnamefont{Hernquist}},
  \bibnamefont{and} \bibinfo{author}{\bibfnamefont{D.}~\bibnamefont{Keres}},
  \bibinfo{journal}{Astrophys. J.} \textbf{\bibinfo{volume}{709}},
  \bibinfo{pages}{1138} (\bibinfo{year}{2010}), \eprint{0907.3482}.

\bibitem[{\citenamefont{Abdo et~al.}(2009)}]{Abdo:2009mr}
\bibinfo{author}{\bibfnamefont{A.~A.} \bibnamefont{Abdo}} \bibnamefont{et~al.}
  (\bibinfo{collaboration}{Fermi LAT}), \bibinfo{journal}{Phys. Rev. Lett.}
  \textbf{\bibinfo{volume}{103}}, \bibinfo{pages}{251101}
  (\bibinfo{year}{2009}), \eprint{0912.0973}.

\bibitem[{\citenamefont{Abdo et~al.}(2010)}]{Abdo:2010dk}
\bibinfo{author}{\bibfnamefont{A.~A.} \bibnamefont{Abdo}} \bibnamefont{et~al.}
  (\bibinfo{collaboration}{Fermi-LAT}) (\bibinfo{year}{2010}),
  \eprint{1002.4415}.

\bibitem[{\citenamefont{Cirelli et~al.}(2009)\citenamefont{Cirelli, Panci, and
  Serpico}}]{Cirelli:2009dv}
\bibinfo{author}{\bibfnamefont{M.}~\bibnamefont{Cirelli}},
  \bibinfo{author}{\bibfnamefont{P.}~\bibnamefont{Panci}}, \bibnamefont{and}
  \bibinfo{author}{\bibfnamefont{P.~D.} \bibnamefont{Serpico}}
  (\bibinfo{year}{2009}), \eprint{0912.0663}.

\bibitem[{\citenamefont{Abazajian et~al.}(2010)\citenamefont{Abazajian,
  Agrawal, Chacko, and Kilic}}]{Abazajian:2010sq}
\bibinfo{author}{\bibfnamefont{K.~N.} \bibnamefont{Abazajian}},
  \bibinfo{author}{\bibfnamefont{P.}~\bibnamefont{Agrawal}},
  \bibinfo{author}{\bibfnamefont{Z.}~\bibnamefont{Chacko}}, \bibnamefont{and}
  \bibinfo{author}{\bibfnamefont{C.}~\bibnamefont{Kilic}}
  (\bibinfo{year}{2010}), \eprint{1002.3820}.

\bibitem[{\citenamefont{{H{\"u}tsi} et~al.}(2010)\citenamefont{{H{\"u}tsi},
  {Hektor}, and {Raidal}}}]{Hutsi:2010}
\bibinfo{author}{\bibfnamefont{G.}~\bibnamefont{{H{\"u}tsi}}},
  \bibinfo{author}{\bibfnamefont{A.}~\bibnamefont{{Hektor}}}, \bibnamefont{and}
  \bibinfo{author}{\bibfnamefont{M.}~\bibnamefont{{Raidal}}},
  \bibinfo{journal}{Journal of Cosmology and Astro-Particle Physics}
  \textbf{\bibinfo{volume}{7}}, \bibinfo{pages}{8} (\bibinfo{year}{2010}),
  \eprint{1004.2036}.

\bibitem[{\citenamefont{Ando and Komatsu}(2006)}]{Ando:2005xg}
\bibinfo{author}{\bibfnamefont{S.}~\bibnamefont{Ando}} \bibnamefont{and}
  \bibinfo{author}{\bibfnamefont{E.}~\bibnamefont{Komatsu}},
  \bibinfo{journal}{Phys. Rev.} \textbf{\bibinfo{volume}{D73}},
  \bibinfo{pages}{023521} (\bibinfo{year}{2006}), \eprint{astro-ph/0512217}.

\bibitem[{\citenamefont{Ando et~al.}(2007{\natexlab{a}})\citenamefont{Ando,
  Komatsu, Narumoto, and Totani}}]{Ando:2006mt}
\bibinfo{author}{\bibfnamefont{S.}~\bibnamefont{Ando}},
  \bibinfo{author}{\bibfnamefont{E.}~\bibnamefont{Komatsu}},
  \bibinfo{author}{\bibfnamefont{T.}~\bibnamefont{Narumoto}}, \bibnamefont{and}
  \bibinfo{author}{\bibfnamefont{T.}~\bibnamefont{Totani}},
  \bibinfo{journal}{Mon. Not. Roy. Astron. Soc.}
  \textbf{\bibinfo{volume}{376}}, \bibinfo{pages}{1635}
  (\bibinfo{year}{2007}{\natexlab{a}}), \eprint{astro-ph/0610155}.

\bibitem[{\citenamefont{Ando et~al.}(2007{\natexlab{b}})\citenamefont{Ando,
  Komatsu, Narumoto, and Totani}}]{Ando:2006cr}
\bibinfo{author}{\bibfnamefont{S.}~\bibnamefont{Ando}},
  \bibinfo{author}{\bibfnamefont{E.}~\bibnamefont{Komatsu}},
  \bibinfo{author}{\bibfnamefont{T.}~\bibnamefont{Narumoto}}, \bibnamefont{and}
  \bibinfo{author}{\bibfnamefont{T.}~\bibnamefont{Totani}},
  \bibinfo{journal}{Phys. Rev.} \textbf{\bibinfo{volume}{D75}},
  \bibinfo{pages}{063519} (\bibinfo{year}{2007}{\natexlab{b}}),
  \eprint{astro-ph/0612467}.

\bibitem[{\citenamefont{Siegal-Gaskins}(2008)}]{SiegalGaskins:2008ge}
\bibinfo{author}{\bibfnamefont{J.~M.} \bibnamefont{Siegal-Gaskins}},
  \bibinfo{journal}{JCAP} \textbf{\bibinfo{volume}{0810}}, \bibinfo{pages}{040}
  (\bibinfo{year}{2008}), \eprint{0807.1328}.

\bibitem[{\citenamefont{Hensley et~al.}(2009)\citenamefont{Hensley,
  Siegal-Gaskins, and Pavlidou}}]{Hensley:2009gh}
\bibinfo{author}{\bibfnamefont{B.~S.} \bibnamefont{Hensley}},
  \bibinfo{author}{\bibfnamefont{J.~M.} \bibnamefont{Siegal-Gaskins}},
  \bibnamefont{and} \bibinfo{author}{\bibfnamefont{V.}~\bibnamefont{Pavlidou}}
  (\bibinfo{year}{2009}), \eprint{0912.1854}.

\end{thebibliography}

\end{document}